\documentclass[aps,prd,twocolumn,amsmath,amssymb,nofootinbib,superscriptaddress]{revtex4-2}

\usepackage{graphicx}  % include graphics
\usepackage[]{hyperref}  % hyperlinks
\usepackage{physics}   % physics macros (bra-ket, derivatives, etc.)
\usepackage{xcolor}    % colored text
\usepackage{siunitx}
\usepackage{mathtools}
\usepackage{cancel}
\usepackage{placeins}
\usepackage{bbm}

\newcommand{\mV}{\mathcal{V}}

\newcommand{\bs}{\boldsymbol}

\newcommand{\br}[1]{\left\langle#1\right|}
\newcommand{\ke}[1]{\left|#1\right\rangle}

\newcommand{\commentthis}[1]{}

\definecolor{citecol}{rgb}{0,0,0.65}
\hypersetup{colorlinks=true,allcolors=citecol}

\usepackage{enumerate}

\begin{document}
\title{Atomic Observables Induced by Cosmic Fields} 

\author{S. Lahs}
\email{sebastian.lahs@cnrs.fr}
\author{D. Comparat}
\affiliation{Universit\'e Paris-Saclay, CNRS,  Laboratoire Aim\'e Cotton, 91405, Orsay, France}
\author{F. Kirk}
\affiliation{Physikalisch-Technische Bundesanstalt, Bundesallee 100, 38116 Braunschweig, Germany}
\affiliation{Institut für Theoretische Physik, Leibniz Universität Hannover, Appelstraße 2, 30167 Hannover, Germany}
\affiliation{Department of Particle Physics and Astrophysics, Weizmann Institute of Science, Rehovot, Israel 7610001}
\author{B. Roberts}
\affiliation{School of Mathematics and Physics, The University of Queensland, Brisbane QLD 4072, Australia}

\date{\today} % REMEBER: Replace with manual date (or remove) for arXiv (since arXiv randomly recompiles sometimes, with makes this look odd!)

\begin{abstract}
\noindent The existence of cosmic fields made from yet unknown light bosons is predicted in many extensions to the Standard Model. They are especially of interest as possible constituents of dark matter. To detect such light and weakly interacting fields, atomic precision measurements offer one of the most sensitive platforms. In this work, we derive which atomic observables are sensitive to what kind of cosmic field couplings. For this we consider fields that couple either through scalar, pseudoscalar, vector, axial vector, or tensor couplings. We derive the corresponding non relativistic atomic potentials. Based on their symmetry properties, these can induce direct energy shifts or induce atomic electric dipole, magnetic dipole, electric quadrupole as well as nuclear Schiff and anapole moments. 
%\FK{I still think a more specific title would be better since more informative}\SL{I like it like this. What do you suggest?}
\end{abstract}
\maketitle

\section{Introduction}
To answer some of the most pressing questions in the fundaments of physics, the existence of yet unobserved ultralight bosonic fields has been proposed. Such fields could be the components of dark matter \cite{preskill1983cosmology} and dark energy \cite{ratra1988cosmological}. They could offer solutions to the hierarchy \cite{graham2015cosmological} and the strong $CP$ problem \cite{weinberg1978new, wilczek1978problem, kim1979weak}. These types of fields also arise generically in theories explaining the unification of the fundamental forces \cite{fritzsch1975unified, witten1980neutrino}, in string theories \cite{witten1984some,svrcek2006axions, goodsell2009naturally}, in theories with additional hidden spacetime dimensions \cite{witten1981search} or violations of Lorentz invariance \cite{kostelecky1999nonrelativistic}. Evidently, there is a strong motivation to perform experiments searching for such fields. Here, low-energy precision measurements can offer some of the highest levels of precision~\cite{safronova2023searches, o2025cajohare}. Because the landscape of theories predicting them is so vast, it is useful to approach the subject in a way that is mostly agnostic to the high-energy origins of the bosonic fields and to focus instead on the phenomenological properties that can be constrained in experiments; namely, the mass of the boson and how it may couple to regular matter.

In this paper, we expand on previous works \cite{roberts2014limiting, roberts2014parity, gaul2020chiral, berlin2024physical, smith2024fermionic} and derive how the coupling of such a general cosmic field to electrons, protons, and neutrons induces atomic line-shifts, as well as electric dipole, magnetic dipole, electric quadrupole, and nuclear Schiff and anapole moments. The goal in this treatment is 
%not to propose a specific new experiment, but instead 
to show generally which type of experiment can be used to constrain what type of cosmic field interaction. This can help to widen the scope of models that can be tested with existing and upcoming experiments.

We start the derivation in Sec.~\ref{sec: Interaction Lagrangians} by stating the interaction Lagrangians between a fermionic and a general bosonic tensor field, and continue in Sec.~\ref{sec: cosmic fields} with a  discussion of the different types of cosmic fields that we consider in this treatment.
In Sec.~\ref{sec: Atomic interaction potentials}, we find the low-energy atomic operators implied by the interaction Lagrangian, and discuss their most important properties.
In Sec.~\ref{sec: Induced observables} we derive the induced atomic observables, including energy shifts (Sec.~\ref{sec: energy shift}),  electric and magnetic dipole moments (Sec.~\ref{sec: The induced electric dipole moment}, \ref{sec: Induced magnetic dipole and ...}), and nuclear moments that lead to modifications in the atomic hyperfine structure (Secs.~\ref{sec: Induced nuclear moments}, \ref{sec: The nuclear anapole moment}).
%We further show how a cosmic field interaction with protons and neutrons induces nuclear moments that lead to modifications in the atomic hyperfine structure (Sec.~\ref{sec: Induced nuclear moments}, \ref{sec: The nuclear anapole moment}).

While we focus specifically on atoms, the results apply equally for ions and (small) molecules.
Throughout this paper, we employ natural units: $c=\hbar=1$, and use the Dirac representation for Dirac matrices.
We use Einstein's summation convention, with Greek and Latin indices running from $0$ to $3$ and $1$ to $3$, respectively, and use the metric of negative signature.

\section{Interaction Lagrangians} \label{sec: Interaction Lagrangians}
In order to cover a wide range of possible atomic observables, we should start by considering how a cosmic field could couple to an atomic system in the first place. The field might either interact with its electrons or nucleons. The most general current of such spin-1/2 fermions $\psi$ is given by $\bar{\psi}\Gamma_{\mu\nu}\psi$. Here, $\Gamma_{\mu\nu}$ is an arbitrary constant $4\times4$ matrix. If there exists some arbitrary cosmic field $\Xi^{\mu\nu}$ that couples through a renormalizable nonderivative coupling with strength $g_\Xi$, then the interaction is simply given by:
\begin{align}
  \mathcal{L}_{\Xi}=- g_\Xi\,\bar{\psi}\,\Gamma\Xi\,\psi\label{eq: LXi}
\end{align}
with $\Gamma \in \{\mathbbm{1}, i\gamma^5,\gamma_\mu, \gamma_\mu\gamma^5 , \sigma_{\mu\nu}\}$, where $\sigma_{\mu\nu}=\tfrac{i}{2}\big[\gamma_\mu,\gamma_\nu\big]$ \cite{peskin2018introduction}.
%\FK{Ben, can you add extra motivation from the atomic side why these currents are interesting and sufficient?} 
This decomposition into Lorentz structures allows us to distinguish 5 Lagrangians $\mathcal{L}_\Xi$:
\begin{align}
\begin{split}
  \mathcal{L}_\phi=&-g_\phi\bar{\psi} \phi \psi, 
  \qquad\quad~~ \mathcal{L}_a=-ig_a \bar{\psi} a \gamma^5 \psi,\\
  \mathcal{L}_{A'}=&-g_{A'} \bar{\psi} \gamma^\mu A'_\mu \psi\,, \quad\mathcal{L}_{Z'}=- g_{Z'}\bar{\psi}\gamma^\mu Z'_\mu \gamma^5 \psi, \\
  \mathcal{L}_\Theta=&-g_{\Theta}\bar{\psi}\sigma_{\mu\nu}\Theta^{\mu\nu} \psi,
\end{split}
\end{align}
%\FK{why minus signs? E.g. Ref.~\cite{frugiuele2022muonic} does not have minus signs. Check difference $\mathcal{L}$ vs $\mathcal{H}$?}\SL{It is just a choice because we wont know the sign of the coupling constants. But because it is T-V in the Lagrangian, I decided to put it with a minus. Changing the sign here, will essentially change it everywhere.}
where $\phi$ is a scalar, $a$ a pseudoscalar, $A'_\mu$ a vector, $Z'_\mu$ an axial vector, and $\Theta_{\mu\nu}$ a tensor field, and $g_\phi$, $g_a$, $g_{A'}$, $g_{Z'}$, $g_\Theta$ are the respective coupling constants between the fermion $\psi$ and the respective field.

Like the electromagnetic four-potential, the vector fields $A'_\mu$, and $Z'_\mu$ consist of scalar ($A'_0$, $Z'_0$) and 3D vector ($\bs{A}'$, $\bs{Z}'$) quantities. For the tensor field $\Theta_{\mu\nu}$, only antisymmetric components contribute to the interaction (see Appendix \ref{sec: tensor coupling}). In analogy to the electromagnetic field tensor, we define $\bs{\theta}^{E,B}$ via:
\begin{equation}
    \Theta^{\mu\nu}=
      \begin{pmatrix}
        0 & -\theta_x^E & -\theta_y^E & -\theta_z^E \\
        \theta_x^E & 0 & -\theta_z^B & \theta_y^B \\
        \theta_y^E & \theta_z^B & 0 & -\theta_x^B \\
        \theta_z^E & -\theta_y^B & \theta_x^B & 0
      \end{pmatrix}.
\end{equation}

\section{Cosmic Fields} \label{sec: cosmic fields}

\begin{figure}%[t!]
    \includegraphics[width=0.3\textwidth]{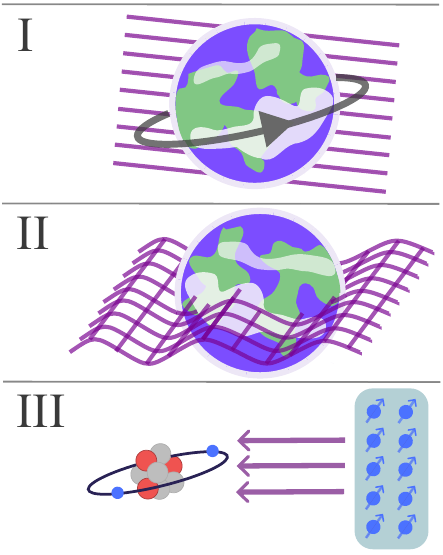}
    \caption{Illustration of the different types of Cosmic fields discussed here. Type I fields are completely static on terrestrial scales. They can be observed through their polarizations that also change over time due to the relative movement between the lab and the reference frame the field is fixed in. Type II fields are plane waves on the length scales of the experiment. Type III fields are sourced from local macroscopic test masses.}
    \label{fig:cosmic field}
\end{figure}

\textit{Cosmic fields} is an umbrella term to describe any new physics field that extends over large enough length-scales for it to appear homogeneous on the scale of the experiment. It is therefore necessarily external to the atomic system. We further assume that the cosmic fields has such a high occupation number that it can be approximated as a classical field. 
%In many aspects, the interaction with cosmic fields $\Xi^{\mu\nu}$ is quite analogous to the semiclassical interaction with the electromagnetic field $F^{\mu\nu}$. 
We can distinguish between three phenomenological types of cosmic fields, as illustrated in Fig.~\ref{fig:cosmic field}.
Even though they have quite different properties, we will be able to treat all within the same formalism:

\paragraph*{Type I:} The mathematically simplest version of a cosmic field $\Xi^{\mu\nu}$ is one that is defined by a constant matrix with no spatial or temporal dependencies (similar to a static homogeneous electric/magnetic field). Such static fields exist, for example, in the Standard Model Extension (SME)~\cite{kostelecky1999nonrelativistic}. In this framework, $\Xi^{\mu\nu}$ would quantify the magnitude and type of (apparent) violations of Lorentz symmetry.

\paragraph*{Type II:}  Most of the theories mentioned in the introduction predict the existence of new bosons. If these would have been produced in large amounts in the early stages of the universe and have only feeble interactions with regular matter, they could have survived in large numbers to the present day. If such bosonic fields further have weak self-interactions and light (but non-zero) masses, they would form a gravitationally bound condensate in every galaxy and therefore also around Earth. 

On the length scale of atoms, these fields take the form of plane waves. For example, the scalar field is given by:
\begin{align}
 \phi(\bs{r},t)&=\mathrm{A}_\phi\mathrm{Re} \Big[\exp{(im_{\phi}\bs{v}\cdot{\bs{r}}-im_{\phi} t)}\Big]\label{eq: cosmic fields wave eq},
\end{align}
where $\mathrm{A}_\phi$ is the amplitude of the field, $m_\phi$ is the boson's mass, and $\bs{v}$ is the relative velocity between the lab frame and the rest frame of the field. The fields $a$, $A_0'$, and $Z'_0$ follow analogously. 

The 3D vector fields $\bs{A}'$ $\bs{Z}'$, $\bs{\theta}^E$, and $\bs{\theta}^B$ include an additional polarization vector $\bs{\xi}$.  For example:
 \begin{align}
 \bs{A}'(\bs{r},t)&= \mathrm{A}_{A'}\mathrm{Re} \Big[\bs{\xi}^{A'} \,\exp{(im_{A'}\bs{v}\cdot{\bs{r}}-im_{A'} t)}\Big]\label{eq: cosmic fields wave eq 2}.
 \end{align}
While the wave vector of an electromagnetic field is necessarily orthogonal to its polarization, this is not true for massive fields, which possess three polarization directions.

Probably the most extensively studied example of a cosmic field is that of axion dark matter. The QCD axion is a pseudoscalar field $a$ and was originally introduced to solve the strong CP problem \cite{Wilczek:1977pj,weinberg1978new,wilczek1978problem, preskill1983cosmology, kim2010axions}. Axion-like particles also appear in many other theories, such as in the form of the relaxion that offers a solution to the hierarchy problem \cite{graham2015cosmological} and as a generic feature of string theories \cite{witten1984some,svrcek2006axions, goodsell2009naturally}. In Refs.~\cite{graham2015experimental,o2025cajohare}, an overview of the various experiments that search for axions and axion-like particles can be found.

Scalar fields $\phi$ appear for example in models of modified gravity \cite{arkani2004ghost, capozziello2011extended}, in dark energy models such as quintessence and chameleon fields \cite{ratra1988cosmological, olive2008environmental, linder2008dynamics, tsujikawa2013quintessence}, and theories with extra dimensions\cite{witten1981search, jackson2023search}. They have also been considered as candidates for dark matter \cite{ferreira2021ultra}. One common way to search for these fields is by atomic clock comparisons \cite{arvanitaki2015searching, safronova2019search}.

Another type of cosmic field are so-called \textit{dark photons}. Such particles can arise through additional $U(1)$ symmetry groups that are predicted in theories of grand unification \cite{fritzsch1975unified, witten1980neutrino}, but also in theories with additional hidden dimensions and string theories \cite{cheng2002kaluza, abel2008kinetic, goodsell2009naturally, jackson2023search}. Dark photons could influence atoms through vector $A'$, axial vector $Z'$, and tensor $\Theta$ interactions. As introduced above, these fields can interact both through scalar and 3D vector components. Whether these are polarized or unpolarized depends on their production mechanisms and self-interactions, which are strongly model dependent \cite{caputo2021dark}. 

It is possible that electrons and nucleons are not directly charged under the new (approximate) $U(1)$ symmetry group. In this case, they still might be able to interact with them through an anomalous dipole coupling. 
Similar to how the uncharged neutron can interact with the electromagnetic field through the anomalous magnetic Pauli moment $\mathcal{L}_\text{Pauli}=-\kappa \frac{\mu_B}{2}\bar{\psi}\sigma_{\mu\nu} F^{\mu\nu}\psi$ \cite{bethe2012quantum}, the interaction with the dark photon cosmic field could take the form $\mathcal{L}_\Theta=-g_{\Theta}\bar{\psi}\sigma_{\mu\nu}\Theta^{\mu\nu} \psi$. 

Both type I and type II cosmic fields can obtain an additional time dependency coming from the relative movement of the lab frame with respect to the rest frame of the field. If there is, for example, a field with a polarization direction fixed with respect to the cosmic microwave background, the sidereal rotation of the Earth will make the polarization time-dependent in the lab frame \cite{kostelecky2015lorentz} (see Fig.~\ref{fig:cosmic field}, I).

\paragraph*{Type III:}  Apart from the static and oscillating cosmic fields discussed above, a third possibility is that new bosonic fields exist not as remnants from the beginning of the universe, but instead appear as new forces (so-called 5th forces) between regular matter particles \cite{Dobrescu:2006au,cong2025spin}. If such forces are sourced by large test masses in the laboratory, they also appear homogeneous on the scale of an atomic system. This is analogous to how one can use capacitors and permanent magnets to generate electric and magnetic fields in the laboratory. Spatial and time dependencies of 5th force potentials can be created by moving the test masses or changing their spin polarizations \cite{terrano2015short}.
A comprehensive overview of all the different types of 5th force interactions and their underlying motivations was recently presented in Ref.~\cite{cong2025spin}.

% This list of examples of cosmic fields is far from complete, but it should give an idea of why experimental searches for them are well-motivated. 

\section{Atomic Interaction Potentials}\label{sec: Atomic interaction potentials}
The Lagrangian to describe a fermion coupled to the cosmic field is given by:
\begin{align}
      \mathcal{L}=i\bar{\psi}\gamma^\mu\partial_\mu\psi-m\bar{\psi}\psi-g_{\Xi}\bar{\psi}\Gamma_{\mu\nu}\Xi^{\mu\nu}\psi
\end{align}
Through the use of the Euler-Lagrange equation, it leads to the following relativistic Hamiltonian:
\begin{align*}
 H\psi=(H^0+V_{\Xi})\psi
 =i\partial_t\psi
 =\big(\bs{\alpha}\cdot\bs{p}+\beta m+\beta\Gamma_{\mu\nu}\Xi^{\mu\nu}\big)\psi
\end{align*}
Here, $\bs{\alpha}=\gamma^0\bs{\gamma}$ and $\beta=\gamma^0$ are the Dirac matrices. % from the Dirac equation (See Appendix \ref{sec: Gamma Matrices}). 
We want to describe a semiclassical interaction between an Atom and the cosmic field. In this context, $\psi$ describes the atomic wavefunction and $\Xi^{\mu\nu}(\bs{r},t)$ becomes an operator which acts on $\psi$. In a form like this, the expression is not very insightful yet. For this reason, we derive in Appendix \ref{sec: Low energy limits of the cosmic field interactions} nonrelativistic interaction potentials. While these may be less accurate for heavy atoms, the resulting potentials are more transparent and help us to connect the different types of interactions with experimental observables. The nonrelativistic limit follows by treating the rest mass of the electron (proton, neutron) as the dominant energy scale and then developing the equation in powers of $m^{-1}$. For the scalar $\psi$, pseudoscalar $a$, vector $V$, axial vector $A$, and tensor $\Theta$ couplings, this yields the following potentials:
\begin{widetext}
\begin{align}
    V_\phi=g_\phi&\Bigg[-\frac{1}{4m^2}\phi \,p^2-\frac{i}{4m^2}(\bs{\nabla}\phi)\cdot\bs{\sigma}(\bs{\sigma}\cdot\bs{p})\Bigg]\label{eq: all potentials}\\[8pt]
    V_a=g_a&\Bigg[-\frac{1}{2m}\bs{\sigma}\cdot(\bs{\nabla}{a})+\frac{i}{4m^2}\bs{\sigma}\cdot(\bs{\nabla}\dot{a})+\frac{1}{4m^2}\dot{a}(\bs{\sigma\cdot\bs{p}})\Bigg]\label{eq: all potentials a}
\end{align}
\begin{align}
    V_{A'}=g_{A'}&\Bigg[\frac{1}{4m^2}A'_0\,p^2-\frac{1}{m}\bs{A}'\cdot\bs{p}-\frac{1}{2m}(\bs{\nabla}\times\bs{A}')\cdot\bs{\sigma}-\frac{i}{4m^2}(\bs{\nabla}A'_0)\cdot\bs{\sigma}(\bs{\sigma\cdot\bs{p}})\label{eq: all potentials A}
    \\
    +&\frac{i}{4m^2}(\bs{\nabla}\times\dot{\bs{A}}')\cdot\bs{\sigma}
    +\frac{i}{4m^2}\dot{\bs{A}}'\cdot\bs{p}-\frac{1}{4m^2}\dot{\bs{A}}'\cdot(\bs{\sigma}\times\bs{p})\Bigg]\nonumber
\end{align}
\begin{align}
    V_{Z'} = g_{Z'}&\Bigg[\bs{\sigma}\cdot\bs{Z}'-\frac{1}{m}Z'_0(\bs{\sigma}\cdot\bs{p})+\frac{i}{2m}(\bs{\nabla}Z'_0)\cdot\bs{\sigma}-\frac{i}{4m^2}\dot{Z}'_0(\bs{\sigma\cdot\bs{p}})\label{eq: all potentials Z}\\
    -&\frac{1}{4m^2}(\bs{\nabla}\dot{Z}'_0)\cdot\bs{\sigma}-\frac{i}{4m^2}(\bs{\nabla}\cdot\bs{Z}')(\bs{\sigma}\cdot\bs{p})\nonumber\\
    +&\frac{1}{4m^2}(\bs{\nabla}\times\bs{Z}')\cdot\bs{\sigma}(\bs{\sigma}\cdot\bs{p})+\frac{1}{2m^2}\bs{Z}'\cdot\bs{p}(\bs{\sigma\cdot\bs{p}})\Bigg]\nonumber
\end{align}
\begin{align}
    V_\Theta = g_\Theta&\Bigg[-2\bs{\sigma}\cdot\bs{\theta}^B-\frac{i}{m}(\bs{\nabla}\times\bs{\theta}^E)\cdot\bs{\sigma}+\frac{2}{m}\bs{\theta}^E\cdot(\bs{\sigma}\times\bs{p})\label{eq: all potentials Th}\\
    -&\frac{1}{2m^2}(\bs{\nabla}\times\dot{\bs{\theta}}^E)\cdot\bs{\sigma}-\frac{1}{2m^2}\dot{\bs{\theta}}^E\cdot\bs{p}-\frac{i}{2m^2}\dot{\bs{\theta}}^E\cdot(\bs{\sigma}\times\bs{p})\nonumber\\
    +&\frac{i}{2m^2}(\bs{\nabla}\cdot\bs{\theta}^B)(\bs{\sigma}\cdot\bs{p})-\frac{1}{2m^2}(\bs{\nabla}\times\bs{\theta}^B)\cdot\bs{\sigma}(\bs{\sigma}\cdot\bs{p})-\frac{1}{m^2}\bs{\theta}^B\cdot\bs{p}(\bs{\sigma}\cdot\bs{p})\Bigg]\nonumber
\end{align}
\end{widetext}
%\FK{widetext doesn't seem necessary here, the formulas are sufficiently short}\SL{I'm ok with both. If you want to put more line breaks to make the equation fit into te single column, you can do it. But make sure to not change the order of the terms as the table lists them in the same order.}
Here, $m$ is the electron (proton, neutron) mass, $\bs{p}=-i\bs{\nabla}$ is the momentum operator, and $\sigma_i$ are the Pauli matrices, and  $\dot{a}$ is the time derivative of the field $a$.

One can see that the terms in Eqs.~(\ref{eq: all potentials}--\ref{eq: all potentials Th}) always consists of an operator that acts on the atomic states, multiplied by a cosmic field or its derivative.
We can summarize the structure as:
\begin{align}
 V_\Xi&=\sum_j V_{\Xi_j}=g_\Xi\sum_j  h^j \mathcal{O}^j f^j(\Xi)
\end{align}
where $g_\Xi$ is the coupling constant, $h^j$ is either $1$ or $i$, $\mathcal{O}^j$ is an Hermitian atomic operator, and $f^j(\Xi)$ is the cosmic field or derivative. 

We now would like to understand how these interactions could be observed in an experiment. In particular, we want to see which of these terms can cause direct energy shifts or induce electric or magnetic dipole moments. In Appendix \ref{app: Derivation of the...}, we derive which properties an atomic operator needs to possess to induce certain observables. What matters is the rank $k$ of $\mathcal{O}^j$ and its transformation behavior under parity $P$ and time reversal $T$. While the whole Hamiltonian of the atom-field interaction is Hermitian and therefore energy conserving, the individual atomic operators can be either Hermitian or anti-Hermitian, and need to be treated accordingly.

We summarize all important properties of the operators from Eqs.~(\ref{eq: all potentials}--\ref{eq: all potentials Th}) in Table \ref{tab:catalog}. We note if it is a scalar ($k=0$) or vector ($k=1$) quantity. Using the fact that $\sigma_i$ is $P$-even, $T$-odd, and $p_i$ is $P$-odd, $T$-odd, we note if the atomic operators are even or odd under $P$ and $T$. 

\begin{table}%[h!]
\centering
\begin{tabular}{|ccc|cccc|}
\hline
Name & Operator $\mathcal{O}$& Field $f(\Xi)$& $k$ & $P$ & $T$ & $h$ \\ \hline
$\phi_1$   & $-\frac{1}{4m^2}p^2$& $\phi$                  & 0  & +  & +  & +  \\
$\phi_2$   & $-\frac{1}{4m^2}\bs{\sigma}(\bs{\sigma}\cdot\bs{p})$& $(\bs{\nabla}\phi)$            & 1  & -  & -  & -  \\\hline
$a_1$     & $-\frac{1}{2m}\bs{\sigma}$& $(\bs{\nabla}a)$             & 1  & +  & -  & +  \\
$a_2$     & $\frac{1}{4m^2}\bs{\sigma}$& $(\bs{\nabla}\dot{a})$          & 1  & +  & -  & -  \\
$a_3$     & $\frac{1}{4m^2}(\bs{\sigma}\cdot\bs{p})$& $\dot{a}$                 & 0  & -  & +  & +  \\\hline
$A'_{1}$   & $\frac{1}{4m^2}p^2$& $A'_0$                  & 0  & +  & +  & +  \\
$A'_{2}$   & $-\frac{1}{m}\bs{p}$& $\bs{A}'$                 & 1  & -  & -  & +  \\
$A'_{3}$ & $-\frac{1}{2m}\bs{\sigma}$& $(\bs{\nabla}\times \bs{A}')$       & 1  & +  & -  & +  \\
$A'_{4}$ & $-\frac{1}{4m^2}\bs{\sigma}(\bs{\sigma}\cdot\bs{p})$& $(\bs{\nabla}A'_0)$            & 1  & -  & -  & -  \\
$A'_{5}$ & $\frac{1}{4m^2}\bs{\sigma}$& $(\bs{\nabla}\times\dot{ \bs{A}}')$    & 1  & +  & -  & -  \\
$A'_{6}$ &$\frac{1}{4m^2}\bs{p}$& $\dot{ \bs{A}}'$             & 1  & -  & -  & -  \\
$A'_{7}$ & $-\frac{1}{4m^2}(\bs{\sigma}\times\bs{p})$& $\dot{ \bs{A}}'$             & 1  & -  & +  & +  \\\hline
$Z'_{1}$   & $\bs{\sigma}$& $\bs{Z}'$                 & 1  & +  & -  & +  \\
$Z'_{2}$   & $\frac{1}{m}(\bs{\sigma}\cdot\bs{p})$& ${Z}'_0$                 & 0  & -  & +  & +  \\
$Z'_{3}$   & $-\frac{1}{2m}\bs{\sigma}$& $(\bs{\nabla}Z'_0)$            & 1  & +  & -  & -  \\
$Z'_{4}$   & $-\frac{1}{4m^2}(\bs{\sigma}\cdot\bs{p})$& $\dot{Z}'_0$               & 0  & -  & +  & -  \\
$Z'_{5}$   & $-\frac{1}{4m^2}\bs{\sigma}$& $(\bs{\nabla}\dot{Z}'_0)$         & 1  & +  & -  & +  \\
$Z'_{6}$   & $-\frac{1}{4m^2}(\bs{\sigma}\cdot\bs{p})$& $(\bs{\nabla}\cdot \bs{Z}')$       & 0  & -  & +  & -  \\
$Z'_{7}$   & $\frac{1}{4m^2}\bs{\sigma}(\bs{\sigma}\cdot\bs{p})$& $(\bs{\nabla}\times \bs{Z}')$       & 1  & -  & -  & +  \\
$Z'_8$    & $\frac{1}{2m^2}\bs{p}(\bs{\sigma}\cdot\bs{p})$& $\bs{Z}'$                 & 1  & +  & -  & +  \\\hline
$\Theta_{1}$ & $-2\bs{ \sigma}$& $\bs{\theta}^B$              & 1  & +  & -  & +  \\
$\Theta_{2}$ & $-\frac{1}{m}\bs{\sigma}$& $(\bs{\nabla}\times \bs{\theta}^E)$    & 1  & +  & -& -  \\
$\Theta_{3}$ & $\frac{2}{m}(\bs{\sigma}\times\bs{p})$& $\bs{\theta}^E$              & 1  & -  & +  & +  \\
$\Theta_{4}$ & $-\frac{1}{2m^2}\bs{\sigma}$& $(\bs{\nabla}\times\dot{ \bs{\theta}}^E)$ & 1  & +  & -  & +  \\
$\Theta_{5}$ & $-\frac{1}{2m^2}\bs{p}$& $\dot{ \bs{\theta}}^E$          & 1  & -  & -  & +  \\
$\Theta_{6}$ & $-\frac{1}{2m^2}(\bs{\sigma}\times\bs{p})$& $\dot{ \bs{\theta}}^E$          & 1  & -  & +  & -  \\
$\Theta_{7}$ & $\frac{1}{2m^2}(\bs{\sigma}\cdot\bs{p})$& $(\bs{\nabla}\cdot \bs{\theta}^B)$    & 0  & -  & +  & -  \\
$\Theta_8$ & $-\frac{1}{2m^2}\bs{\sigma}(\bs{\sigma}\cdot\bs{p})$& $(\bs{\nabla}\times \bs{\theta}^B)$    & 1  & -  & -  & +  \\
$\Theta_9$ & $-\frac{1}{m^2}\bs{p}(\bs{\sigma}\cdot\bs{p})$& $\bs{\theta}^B$              & 1  & +  & -  & +  \\ \hline
\end{tabular}
 \caption{The atomic operators $\mathcal{O}$ and the cosmic field $f({\Xi})$ they couple to [see Eqs.~(\ref{eq: all potentials}--\ref{eq: all potentials Th})]. We give the rank $k$ of the operators and their transformation behaviour under parity $P$ and time reversal $T$. Finally, we use $h$ to indicate if the operator appears with (-) or without (+) a prefactor of $i$.}
  \label{tab:catalog}
\end{table}
% \FloatBarrier

\subsection{Interpretation of the Interaction Potentials} \label{sec: interpretation of the atomic potentials}
In table \ref{tab:catalog}, 7 different types of atomic operators $\mathcal{O}$ appear. In the following, we want to discuss their properties to get an idea what atomic observables they might lead to:
\begin{enumerate}[(i)]
\item  $p^2$: The terms $\phi_1$ and $A'_1$ depend on the scalar operator $p^2$. They therefore describe a modification of the kinetic energy of the atom. Alternatively, this effect can also be understood as a modification of the electron (proton, neutron) mass: 
\begin{align}
    \frac{p^2}{2m}+V_{\phi_1}+V_{A'_1}&=\frac{p^2}{2}\bigg[\frac{1}{m}-\frac{1}{m^2}\Big[\tfrac{1}{2}g_\phi\phi-\tfrac{1}{2}g_{A'}A'\Big]\bigg]\nonumber\\
    &\simeq\frac{p^2}{2\big[m+m'(t)\big]}
\end{align}

\item $(\bs{\sigma}\cdot\bs{p})$: The terms $a_3$, $Z'_2$, $Z'_4$, $Z'_6$, and $\theta_7$ couple the cosmic fields to the scalar operator $(\bs{\sigma}\cdot\bs{p})$. This quantity is also known as the chiral charge and is connected to the low energy limit of $\gamma_5$. 
This indicates that these interactions induce (apparent) atomic parity violation. 

\item $\bs{\sigma}$: The operators in terms $a_1$, $a_2$, $A'_3$, $A'_5$, $Z'_1$, $Z'_3$, $Z'_5$, $\Theta_1$, $\Theta_2$ and $\Theta_4$ are all directly proportional to the electron spin $\bs{\sigma}$. They therefore induce an interaction reminiscent of the Zeeman coupling and the cosmic fields they couple to appear as pseudo-magnetic fields. The dependency on $\bs{\sigma}$ makes paramagnetic $(J>0)$ systems best suited to search for these types of interactions.

\item $\bs{p}$: The terms $A'_2$, $A'_6$, and $\Theta_5$ have atomic operators that are directly proportional to the momentum operator $\bs{p}$. The interaction takes the same form as the classical interaction with the electromagnetic vector potential. The fields of these operators, therefore, act as pseudo-electric fields and slightly polarize the atom.

\item $(\bs{\sigma}\times \bs{p})$: The terms, $A'_7$, $\Theta_3$, and $\Theta_6$ involve the vector operator $(\bs{\sigma}\times \bs{p})$. Such an operator also appears classically in the relativistic corrections to the electron-electric field coupling \cite{bethe2012quantum}. It can therefore be viewed as another, more exotic coupling to pseudo-electric fields.

The remaining terms contain atomic operators constructed from multiplying the operators of cases 3) and 4) with the chiral charge from case 2). The terms, therefore, appear as simultaneous atomic parity violation and interactions with pseudo-electromagnetic fields. In particular, we have:

\item $\bs{\sigma}(\bs{\sigma}\cdot\bs{p})$: Terms $\phi_2$, $A'_4$, $Z'_7$ and $\Theta_8$ involve the operator $\bs{\sigma}(\bs{\sigma}\cdot\bs{p})$ that describes a chiral coupling to a pseudo-magnetic field. Through the algebraic relations of the Pauli matrices, it can also be expressed as $\bs{\sigma}(\bs{\sigma}\cdot\bs{p})=\bs{p}-i(\bs{\sigma}\times\bs{p})$. One can therefore also understand this operator as a coupling to pseudo-electric fields.

\item  $\bs{p}(\bs{\sigma}\cdot\bs{p})$: Finally, terms $Z'_8$ and $\Theta_9$ involve the operator $\bs{p}(\bs{\sigma}\cdot\bs{p})$  that describes a chiral version of the electric dipole coupling. The operator can alternatively be written as $-\sigma_j\partial_i\partial_j$, effectively coupling the spin and field to the Hessian of the wavefunction.
\end{enumerate}

Terms $A'_2$, $A'_3$, $A'_5$, $A'_6$, $A'_7$, $Z'_1$, $Z'_6$, $Z'_7$, $Z'_8$, and all $\Theta$ terms depend on the polarization of the bosonic field. If the cosmic field is unpolarized, these terms will be very difficult to detect, as most observables will average to zero.

Terms $\phi_2$, $a_1$, $a_2$, $A'_4$, $Z'_3$, $Z'_5$ depend on the gradient, terms $Z'_6$, $\Theta_7$ depend on the divergence, and terms $A'_3$, $A'_5$, $Z'_7$, $\Theta_2$, $\Theta_4$, $\Theta_8$ on the curl of the cosmic field. For type I cosmic field, these terms will be unobservable. For type II and III, they depend on the relative velocity between the lab frame and the cosmic field [see Eqs.~(\ref{eq: cosmic fields wave eq}), (\ref{eq: cosmic fields wave eq 2})]. This can be thought of as a kind of \textit{Cosmic Wind} that the atomic operator couples to \cite{graham2013new}.

The terms $a_2$, $a_3$, $A'_5$, $A'_6$, $A'_7$, $Z'_4$, $Z'_5$, $\Theta_4$, $\Theta_5$, $\Theta_6$ depend on time derivatives. For type II fields, this leads to a phase shift of $\pi/2$ and a factor of the mass of the boson. For tipe I fields, terms $A'_6$, $A'_9$, $\Theta_5$, $\Theta_6$ would still be observable due to their polarization that is not fixed in the lab frame (see Fig.~\ref{fig:cosmic field}).

Even without considering a specific system, one can get a rough idea of how the size of different terms likely compares to each other. Besides the shared coupling constants $g$, the size of the terms is governed by the suppression factors of the electron mass $m$. Because all operators have the same units (mass dimensions), the terms always contain another quantity that gets compared to $m$. This is either the electron momentum $\bs{p}$, the spatial derivative $\bs{\nabla}$, or the time derivative $\partial_t$. While $\bs{p}$ depends on the atomic system and state, $\bs{\nabla}$, and $\partial_t$ depend only on the cosmic field. If the cosmic field is made up of ultralight bosons, both types of derivatives scale with the mass of the boson ($<10\,$eV \cite{jackson2023search, antypas2022new}). To increase the size of $\bs{p}$, it is preferential to use atoms with high atomic numbers $Z$ \cite{bouchiat1974weak, khriplovich1991parity}.

\section{Induced Observables}\label{sec: Induced observables}
Following the insight that most of the cosmic field interactions are the same as couplings to electric and magnetic fields, there are many different atomic observables one could consider. In the following, we will use the properties from Table~\ref{tab:catalog} and the results from Appendix \ref{app: Derivation of the...} to identify which type of cosmic field interactions will be detectable as direct energy shifts or induced electric and magnetic dipole moments. We further discuss how they could induce atomic electric quadrupole, as well as various nuclear moments.

\subsection{Energy Shifts}\label{sec: energy shift}
The most straightforward way of observing the interaction with the cosmic fields is to measure energy splitting or shifts that affect the atomic spectrum. Only a small subsection of the terms in Eqs.~(\ref{eq: all potentials}--\ref{eq: all potentials Th}) can induce such an energy shift at linear order (see Sec.~\ref{sec: energy shifts}). Only terms with signatures $(0 + + +)$, $(0+--)$, $(1+ - +)$, $(1++-)$ in Table~\ref{tab:catalog} can contribute. The energy shifts due to the different types of couplings are given by:
\begin{widetext}
\begin{align}
  &\Delta E_\phi=g_\phi\langle \mathcal{O}^{\phi_1}\rangle\phi\\[8pt]
   &\Delta E_a=g_a\tfrac{1}{J+1}\langle\mathcal{O}^{a_1}_z\rangle\langle\bs{J}\rangle\cdot(\bs{\nabla}a)\\[8pt]
    &\Delta E_{A'}=g_{A'}\Big[\langle \mathcal{O}^{A'_1}\rangle A'_0+\tfrac{1}{J+1}\langle \mathcal{O}^{A'_3}_z\rangle\langle\bs{J}\rangle\cdot(\bs{\nabla}\times \bs{A}')\Big]\\[8pt]
   &\Delta E_{Z'}=g_{Z'}\Big[\tfrac{1}{J+1}\langle \mathcal{O}^{Z'_1}_z\rangle\langle\bs{J}\rangle\cdot\bs{Z}'+\tfrac{1}{J+1}\langle \mathcal{O}^{Z'_5}_z\rangle\langle\bs{J}\rangle\cdot(\bs{\nabla}\dot{Z}'_0)+\tfrac{1}{J+1}\langle \mathcal{O}^{Z'_8}_z\rangle\langle\bs{J}\rangle\cdot\bs{Z}'\Big]\\[8pt]
   &\Delta E_{\Theta}=g_{\Theta}\Big[\tfrac{1}{J+1}\langle\mathcal{O}^{\Theta_1}_z\rangle\langle\bs{J}\rangle\cdot\bs{\theta}^B+\tfrac{1}{J+1}\langle\mathcal{O}^{\Theta_4}_z\rangle\langle\bs{J}\rangle\cdot(\bs{\nabla}\times\dot{\bs{\theta}}^E)+\tfrac{1}{J+1}\langle\mathcal{O}^{\Theta_9}_z\rangle\langle\bs{J}\rangle\cdot\bs{\theta}^B\Big].
\end{align}
\end{widetext}
Here $\langle \mathcal{O}^{\phi_1} \rangle$ represents the expectation value of the atomic operator of the term $\phi_1$ in the atomic wavefunctions $\psi$. For vector operators like $\bs{\mathcal{O}}^{a_1}$, we can, without loss of generality, take the $z$-component. $J$ is the value of the atom's total angular momentum, and $\langle \bs{J}\rangle$ is the expectation value of the total angular momentum operator. If the nucleus carries a spin, it too will contribute to $J$.

We can see that it is possible for all 5 types of cosmic fields to induce a direct energy shift. In an atomic state with $J=0$, only the $p^2$-dependent terms $\phi_1$ and $A'_1$ contribute. Atomic clock comparisons \cite{arvanitaki2015searching, safronova2019search}, that are usually designed to be insensitive to perturbations through external electromagnetic fields, will mostly be sensitive to exactly these terms. All other terms could be detected through magnetometry-type experiments \cite{afach2024can} as they either depend on $\langle \sigma_z\rangle$ or $\langle p_z(\bs{\sigma}\cdot\bs{p})\rangle$.

Whether terms are detectable or not depends further on the restrictions listed in Sec.~\ref{sec: interpretation of the atomic potentials}. 

\subsection{The Induced Electric Dipole Moment} \label{sec: The induced electric dipole moment}
Apart from relative level shifts, there are more atomic observables that can be measured to high levels of precision - like the electric dipole moment. We will see that atoms can obtain quasi-static and oscillating dipole moments through the interaction with cosmic fields. Mathematically, these effects arise through the perturbation of the electric dipole operator $\bs{d}$ by the atomic operators $\mathcal{O}$. The strength of these interactions can be expressed in the form of generalized polarizabilities $\alpha$. In Appendix \ref{app: Derivation of the...}, we show that all $P$-odd terms contribute to the expectation value of the electric dipole moment $\langle \bs{d}\rangle$. How exactly the polarizabilities couple to external fields depends on their rank, transformation under $T$, and Hermiticity. We find the following induced electric dipole moments:
\begin{widetext}
\begin{align}
  &\langle \bs{d}\rangle_\phi=g_\phi\Big[\alpha^s_{d\phi_2}(\bs{\nabla}{\phi})-\alpha'^v_{d\phi_2}\big[(\bs{\nabla}\dot{\phi})\times\langle\bs{J}\rangle\big]+\alpha^t_{d\phi_2}\mathcal{Q}\cdot(\bs{\nabla}{\phi})\Big]\\[8pt]
 & \langle \bs{d}\rangle_a=-g_a \alpha'_{da_3}\langle\bs{J}\rangle\ddot{a}\\[8pt]
 % \end{align}
 % \begin{align}
  &\langle \bs{d}\rangle_{A'}=g_{A'}\bigg[\Big[-\alpha'^s_{dA'_2}+\alpha^s_{dA'_6}+\alpha^s_{dA'_7}\Big]\dot{\bs{A}}'+\alpha^s_{dA'_4}(\bs{\nabla}A'_0)\\
 &\hspace{33pt}+\alpha^v_{dA'_2}\big[{\bs{A}}'\times\langle\bs{J}\rangle\big]-\Big[\alpha'^v_{dA'_6}+\alpha'^v_{dA'_7}\Big]\big[\ddot{\bs{A}}'\times\langle\bs{J}\rangle\big]-\alpha'^v_{dA'_4}\big[(\bs{\nabla}\dot{A}'_0)\times\langle\bs{J}\rangle\big]\nonumber\\
   &\hspace{33pt}+\Big[-\alpha'^t_{dA'_2}+\alpha^t_{dA'_6}+\alpha^t_{dA'_7}\Big]\mathcal{Q}\cdot\dot{\bs{A}}' +\alpha^t_{dA'_4}\mathcal{Q}\cdot(\bs{\nabla}A'_0)\bigg]\nonumber\\[8pt]
  &\langle \bs{d}\rangle_{Z'}=g_{Z'}\bigg[
   \Big[ - \alpha'_{dZ'_2}+\alpha_{dZ'_4}\Big]\langle\bs{J}\rangle\dot{Z}'_0+\alpha_{dZ'_6}\langle\bs{J}\rangle(\bs{\nabla}\cdot\bs{Z}'_0)-\alpha'^s_{dZ'_7}(\bs{\nabla}\times\dot{\bs{Z}}')\\
   &\hspace{33pt}+\alpha^v_{dZ'_7}\big[(\bs{\nabla}\times{\bs{Z}}')\times\langle\bs{J}\rangle\big]-\alpha'^t_{dZ'_7}\mathcal{Q}\cdot(\bs{\nabla}\times\dot{\bs{Z}}')\bigg]\nonumber%\\[8pt]
 \end{align}
 \begin{align}
   &\langle \bs{d}\rangle_{\Theta}=g_{\Theta}\bigg[+\alpha^s_{d\Theta_3}\bs{\theta}^E-\Big[\alpha'^s_{d\Theta_5}+\alpha'^s_{d\Theta_6}\Big]\ddot{\bs{\theta}}^E+\alpha_{\Theta_7}\langle\bs{J}\rangle(\bs{\nabla}\cdot\bs{\theta}^B)
         -\alpha'^s_{d\Theta_8}(\bs{\nabla}\times\dot{\bs{\theta}}^E)\nonumber\\
   &\hspace{33pt}+\Big[-\alpha'^v_{d\Theta_3}+\alpha^v_{d\Theta_5}+\alpha^v_{d\Theta_6}\Big]\big[\dot{\bs{\theta}}^E\times\langle\bs{J}\rangle\big]+\alpha^v_{d\Theta_8}\big[(\bs{\nabla}\times{\bs{\theta}}^E)\times\langle\bs{J}\rangle\big]\\
   &\hspace{33pt}+\alpha^t_{d\Theta_3}\mathcal{Q}\cdot\bs{\theta}^E-\Big[\alpha'^t_{d\Theta_5}+\alpha'^t_{d\Theta_6}\Big]\mathcal{Q}\cdot\ddot{\bs{\theta}}^E-\alpha'^t_{d\Theta_8}\mathcal{Q}\cdot(\bs{\nabla}\times\dot{\bs{\theta}}^E)\bigg].\nonumber
\end{align} 
\end{widetext}
Here, $
\mathcal{Q}_{ij}\equiv\tfrac{1}{2}\big\langle\big[J_i J_j+J_j J_i-\tfrac{2}{3}\delta_{ij}\sum_lJ_l J_l\big]\big\rangle$ is the angular momentum quadrupole and $\mathcal{Q}\cdot$ describes the matrix product with it. The lower indices of the polarizabilities $\alpha$ indicate which atomic operator they involve, and the upper index identifies which irreducible tensor structure they belong to. The factor $\alpha'^v_{dA'_4}$ for example, is the magnitude of the antisymmetric vector component of the tensor $-2\sum_k\frac{\br{n}d_i\ke{k}\br{k}(V_{A'_4})_j\ke{n}}{\omega_{kn}^2-\omega^2}$ (see Appendix \ref{app: Derivation of the...} and Ref.~\cite{lahs2024polarizabilities}). Here $\omega_{kn}$ is the energy difference between the atomic wavefunctions $\ke{n}$ and $\ke{k}$. $\omega$ is the oscillation frequency of the field. Before accounting for additional effects from the movement of earth, the frequency $\omega$ is zero for type I fields, and equals the mass of the cosmic field $m_\Xi$ for type II fields. For type III fields, a nonzero frequency would be created by shaking the test mass \cite{terrano2015short}.

As described in Sec.~\ref{sec: interpretation of the atomic potentials}, the terms $\phi_2$, $A'_2$, $A'_4$, $A'_6$, $A'_7$, $Z'_7$, $\Theta_3$, $\Theta_5$, $\Theta_6$, $\Theta_8$ are similar to the classical coupling to the electric field. It is therefore not surprising that these interactions can induce electric dipole moments analogous to the Stark effect.

The remaining terms $a_3$, $Z'_2$, $Z'_4$, $Z'_6$, $\Theta_7$ couple to the axial charge $(\bs{\sigma}\cdot\bs{p})$, and through this, induce atomic parity violation. Through either the time derivative of the field or the anti-Hermitian properties of the operator, this parity violation can then induce an electric dipole moment.

Only the terms containing polarizabilities $\alpha^s$ and $\alpha'^s$ are non-vanishing for diamagnetic $(J=0)$ systems. All terms containing $\langle\bs{J}\rangle$ require at least a total angular momentum of $1/2$, while the $\mathcal{Q}$-dependent terms only exist for $J\geq1$. Most of the time, if multiple polarizabilities couple to the same field, only one of them will be dominant for a given atomic system. From electromagnetic polarizabilities, it is known that scalar, vector, and tensor components $\alpha^s$, $\alpha^v$, $\alpha^t$, belonging to the same atomic operator can be of similar size and the induced electric dipole moment can be the result of a complex interplay between them \cite{becher2018anisotropic}. 

For cosmic fields of type II, the terms involving $\alpha$ scale with $\omega_{kn}/(\omega_{kn}^2-m_\Xi^2)$, while the terms with $\alpha'$ scales with $\omega/(\omega_{kn}^2-m_\Xi^2)$. Due to the properties of the electric dipole operator $\bs{d}$, only states$\ke{n}$, $\ke{k}$ that are connected through electric dipole transitions (opposite parity) contribute. If the mass of the cosmic field bosons $m_\Xi$ is sufficiently high, one can optimize the sensitivity by choosing the energy difference $E_n-E_m$ to be close to $m_\Xi$. To detect low-frequency cosmic fields, it is beneficial to work with systems with close-lying parity doublets. If $\omega_{kn}>m_\Xi$, then $\alpha$ terms are amplified compared to $\alpha'$ terms. If $m_\Xi>\omega_{kn}$, it is the other way around.

To detect the electric field like interactions, Rydberg atoms might be a viable system. Through their large electric dipole moments, these systems are very sensitive to fields oscillating in the \SI{}{MHz} to \SI{}{GHz} range \cite{holloway2014broadband, yuan2023quantum}. Similar detection schemes for axions and dark photons have previously been suggested in Refs.~\cite{matsuki1991direct, gue2023search}.

For the other terms, experiments searching for the static electron electric dipole moment (eEDM) could have high sensitivities. In the usual mode of operation, the measurement data are averaged over long periods of time. To deduce limits on cosmic fields, one therefore either needs to reanalyze the timestamped data \cite{abel2017search, roussy2021experimental}, or perform new dedicated searches for such oscillating EDMs \cite{budker2014proposal}. In Ref.~\cite{arvanitaki2024piezoaxionic}, it was suggested to use crystalline environments to couple the oscillating EDM through the piezoelectric effect to the stress tensor of the crystal, which would enable to measure the interaction through the compression and expansion of the solid.

\subsection{Induced Magnetic Dipole and Electric Quadrupole Moments}\label{sec: Induced magnetic dipole and ...}
For both magnetic dipole moment $\bs{\mu}$ and electric quadrupole moment $Q_{ij}$, terms that are $P$-even contribute. The sensitivities are therefore completely complementary to the electric dipole moment. Using again the relations derived in Appendix \ref{app: Derivation of the...}, we obtain the following induced magnetic dipole moments:
\begin{widetext}
\begin{align}
    &\langle \bs{\mu}\rangle_\phi=g_\phi\,\alpha_{\mu \phi_1} \langle\bs{J\rangle}\,\phi \\[8pt]
    &\langle \bs{\mu}\rangle_a=g_a\bigg[\alpha^s_{\mu a_1}(\bs{\nabla}a) -\alpha'^s_{\mu a_2}(\bs{\nabla}\ddot{a})+\Big[-\alpha'^v_{\mu a_1}+\alpha^v_{\mu a_2}\Big]\big[(\bs{\nabla}\dot{a})\times\langle\bs{J}\rangle\big]\\
     &\hspace{33pt}+\alpha^t_{\mu a_1}\mathcal{Q}\cdot(\bs{\nabla}{a})-\alpha'^t_{\mu a_2}\mathcal{Q}\cdot(\bs{\nabla}\ddot{a})\bigg]\nonumber\\[8pt]
   &\langle \bs{\mu}\rangle_{A'}=g_{A'}\bigg[\alpha_{\mu A'_1} A'_0
    +\alpha^s_{\mu A'_3}(\bs{\nabla}\times\bs{A}')-\alpha'^s_{\mu A'_5}(\bs{\nabla}\times\ddot{\bs{A}}')\\
    &\hspace{33pt}+\Big[-\alpha'^v_{\mu A'_3}+\alpha^v_{\mu A'_5}\Big]\big[(\bs{\nabla}\times\dot{\bs{A}}')\times\langle\bs{J}\rangle\big]+\alpha^t_{\mu A'_3}\mathcal{Q}\cdot(\bs{\nabla}\times\bs{A}')-\alpha'^t_{\mu A'_5}\mathcal{Q}\cdot(\bs{\nabla}\times\ddot{\bs{A}}')\bigg]\nonumber\\[8pt]
    &\langle \bs{\mu}\rangle_{Z'}=g_{Z'}\bigg[\Big[\alpha^s_{\mu Z'_1}+\alpha^s_{\mu Z'_8}\Big]\bs{Z}'+\Big[-\alpha'^s_{\mu Z'_3}+\alpha^s_{\mu Z'_5}\Big](\bs{\nabla}\dot{Z}'_0)\\
    &\hspace{33pt} -\Big[\alpha'^v_{\mu Z'_1}+\alpha'^v_{\mu Z'_8}\Big]\big[\dot{\bs{Z}}'\times\langle\bs{J}\rangle\big]+\alpha^v_{\mu Z'_3}\big[(\bs{\nabla}{Z}'_0)\times\langle\bs{J}\rangle\big]-\alpha'^v_{\mu Z'_5}\big[(\bs{\nabla}\ddot{Z}'_0)\times\langle\bs{J}\rangle\big]\nonumber\\
    &\hspace{33pt} +\Big[\alpha^t_{\mu Z'_1}+\alpha^t_{\mu Z'_8}\Big]\mathcal{Q}\cdot\bs{Z}'+\Big[-\alpha'^t_{\mu Z'_3}+\alpha^t_{\mu Z'_5}\Big]\mathcal{Q}\cdot(\bs{\nabla}\dot{Z}'_0)\bigg]\nonumber\\[8pt]
       &\langle \bs{\mu}\rangle_{\Theta}=g_\Theta\bigg[\Big[\alpha^s_{\mu \Theta_1}+\alpha^s_{\mu \Theta_9}\Big]\bs{\theta}^B+\alpha^s_{\mu \Theta_2}(\bs{\nabla}\times\bs{\theta}^E)-\alpha'^s_{\mu \Theta_4}(\bs{\nabla}\times\ddot{\bs{\theta}}^E)\\
       &\hspace{33pt} -\Big[\alpha'^v_{\mu \Theta_1}+\alpha'^v_{\mu \Theta_9}\Big]\big[\dot{\bs{\theta}}^B\times\langle\bs{J}\rangle\big]+\Big[-\alpha'^v_{\mu \Theta_2}+\alpha^v_{\mu \Theta_4}\Big]\big[(\bs{\nabla}\times\dot{\bs{\theta}}^E)\times\langle\bs{J}\rangle\big]\nonumber\\
       &\hspace{33pt}+\Big[\alpha^t_{\mu \Theta_1}+\alpha^t_{\mu \Theta_9}\Big]\mathcal{Q}\cdot\bs{\theta}^B+\alpha^t_{\mu \Theta_2}\mathcal{Q}\cdot(\bs{\nabla}\times\bs{\theta}^E)-\alpha'^t_{\mu \Theta_4}\mathcal{Q}\cdot(\bs{\nabla}\times\ddot{\bs{\theta}}^E)\bigg]\nonumber
\end{align}
For the induced electric quadrupole moments, we get:
\begin{align}
  &\langle Q_{ij}\rangle_\phi=g_\phi\alpha_{Q\phi_1}\mathcal{Q}_{ij}\phi\\[8pt]
  &\langle Q_{ij}\rangle_a=g_a\bigg[\left({\alpha}_{Q a_1}^{t_s}\mathcal{O}_{ l  ij}+{\alpha}_{Q a_1}^{v_m}\mathcal{M}_{ l  ij}\right) (\partial_ l  a)-\left({\alpha'}_{Q a_2}^{t_s}\mathcal{O}_{ l  ij}+{\alpha'}_{Q a_2}^{v_m}\mathcal{M}_{ l  ij}\right) (\partial_ l  \ddot{a})\nonumber\\
  &\hspace{33pt}+\Big[ -{\alpha'}_{Q a_1}^{t_m} +{\alpha}_{Q a_2}^{t_m}\Big]\mathcal{W}_{ l  ij}(\partial_ l \dot{a})\bigg]\\[8pt]
  &\langle Q_{ij}\rangle_{A'}=g_{A'}\bigg[\left({\alpha}_{Q A'_3}^{t_s}\mathcal{O}_{ l  ij}+{\alpha}_{Q A'_3}^{v_m}\mathcal{M}_{ l  ij}\right) (\bs{\nabla}\times\bs{A}')_ l -\left({\alpha'}_{Q A'_5}^{t_s}\mathcal{O}_{ l  ij}+{\alpha'}_{Q A'_5}^{v_m}\mathcal{M}_{ l  ij}\right) (\bs{\nabla}\times\ddot{\bs{A}}')_ l \nonumber\\
   &\hspace{33pt}+\alpha_{QA'_1}\mathcal{Q}_{ij}A'_0+\Big[-{\alpha'}_{Q A'_3}^{t_m}+{\alpha}_{Q A'_5}^{t_m}\Big]\mathcal{W}_{ l  ij}(\bs{\nabla}\times\dot{\bs{A}}')_ l \bigg]\\[8pt]
    &\langle Q_{ij}\rangle_{Z'}=g_{Z'}\bigg[\left(\Big[{\alpha}_{Q Z'_1}^{t_s}+{\alpha}_{Q Z'_8}^{t_s}\Big]\mathcal{O}_{ l  ij}+\Big[{\alpha}_{Q Z'_1}^{v_m}+{\alpha}_{Q Z'_8}^{v_m}\Big]\mathcal{M}_{ l  ij}\right) Z'_ l \\
     &\hspace{33pt}+\left(\Big[-{\alpha}_{Q Z'_3}^{t_s}+{\alpha}_{Q Z'_5}^{t_s}\Big]\mathcal{O}_{ l  ij}+\Big[-{\alpha}_{Q Z'_3}^{v_m}+{\alpha}_{Q Z'_5}^{v_m}\Big]\mathcal{M}_{ l  ij}\right) (\partial_ l  \dot{Z}'_0)\nonumber\\
    &\hspace{33pt}-\Big[{\alpha'}_{Q Z'_1}^{t_m}+{\alpha'}_{Q Z'_8}^{t_m}\Big]\mathcal{W}_{ l  ij}\dot{Z}'_ l  +{\alpha}_{Q Z'_3}^{t_m}\mathcal{W}_{ l  ij} (\partial_ l {Z'_0})-{\alpha'}_{Q Z'_5}^{t_m}\mathcal{W}_{ l  ij} (\partial_ l \ddot{Z}'_0)\bigg]\nonumber\\[8pt]
      &\langle Q_{ij}\rangle_{\Theta}=g_{\Theta}\bigg[\left(\Big[{\alpha}_{Q \Theta_1}^{t_s}+{\alpha}_{Q \Theta_9}^{t_s}\Big]\mathcal{O}_{ l  ij}+\Big[{\alpha}_{Q\Theta_1}^{v_m}+{\alpha}_{Q\Theta_9}^{v_m}\Big]\mathcal{M}_{ l  ij}\right) \theta^B_ l \\
      &\hspace{33pt}+\left({\alpha}_{Q \Theta_2}^{t_s}\mathcal{O}_{ l  ij}+{\alpha}_{Q \Theta_2}^{v_m}\mathcal{M}_{ l  ij}\right) (\bs{\nabla}\times\bs{\theta}^E)_ l -\left({\alpha'}_{Q \Theta_4}^{t_s}\mathcal{O}_{ l  ij}+{\alpha'}_{Q \Theta_4}^{v_m}\mathcal{M}_{ l  ij}\right) (\bs{\nabla}\times\ddot{\bs{\theta}}^E)_ l \nonumber\\
       &\hspace{33pt}-\Big[ {\alpha'}_{Q \Theta_1}^{t_m}+{\alpha'}_{Q \Theta_9}^{t_m}\Big]\mathcal{W}_{ l  ij}\dot{\theta}^B_ l +\Big[-{\alpha'}_{Q \Theta_2}^{t_m}+{\alpha}_{Q \Theta_4}^{t_m}\Big]\mathcal{W}_{ l  ij}(\bs{\nabla}\times\dot{\bs{\theta}}^E)_ l \bigg]\nonumber
\end{align}
\end{widetext}
Here, $\mathcal{O}_{ l  ij}$ is the angular momentum octupole, and $\mathcal{M}_{ l  ij}$, $\mathcal{W}_{ l  ij}$ are other angular momentum structures that are defined in Appendix \ref{app: Derivation of the...}. 

The moments $\langle\bs{\mu}\rangle$ and $\langle Q_{ij}\rangle$ contain all operators that appear in the direct energy shift. Additionally, the two moments are sensitive to the anti-Hermitian terms $a_2$, $A'_5$, $\Theta_2$. Because $a_2$ and $A'_5$ are both suppressed ($\omega/m\ll1$) compared to the terms $a_1$ and $A'_3$, respectively, they are probably not of much interest.

Besides $\phi_1$, $A'_1$ $Z'_8$, and $\Theta_9$ all terms that contribute to $\langle\bs{\mu}\rangle$, and $\langle Q_{ij}\rangle$ have an atomic operator proportional to $\bs{\sigma}$. It is therefore not surprising that they affect the system in the same way as a magnetic field \cite{lahs2024polarizabilities}. 
The $\phi_1$ and $A'_1$ terms induce an oscillating mass component, while $Z'_8$ and $\Theta_9$ depend on the more exotic operator $\bs{p}(\bs{\sigma}\cdot\bs{p})$.

%As it was the case for $\langle \bs{d}\rangle$, neither $\langle \bs{\mu}\rangle$, nor $\langle Q_{ij}\rangle$ contain components that are completely static. It would, anyway, be difficult to distinguish those from the sizable intrinsic magnetic dipole and electric quadrupole moments that atomic systems possess.

The size of the $\alpha$ and $\alpha'$ factors can again be maximized by bringing the energy difference between atomic states into resonance with the oscillation frequency of the cosmic field. This time, only states that are connected by magnetic dipole or electric quadrupole transitions (same parity) will contribute.

Induced magnetic dipole moments can be detected by measuring the magnetization of a large ensemble of atoms. In Ref.~\cite{bloch2023scalar}, it was, for example, suggested to measure the oscillation of the field of a permanent magnet to probe for scalar dark matter. Magnetometry measurements in general can reach exceedingly high precisions and have already been used to put constraints on dark matter interactions \cite{afach2024can}.

Detecting atomic electric quadrupole moments is less straightforward. Because they possess sensitivities to the same terms as magnetic dipole moments, there is also no strong motivation to search for them. Quadrupole moments are more interesting in the context of nuclear moments. We discuss these in the following.

\subsection{Induced Nuclear Moments} \label{sec: Induced nuclear moments}
So far we discussed the consequences of cosmic fields that couple to electrons (leptons). But, depending on the underlying model, the coupling to protons and neutrons (hadrons) could be the dominant way the field interacts with matter. A coupling of the cosmic field to quarks or gluons can be described as an effective coupling to nucleons. This in turn induces nuclear multipole moments. These can through hyperfine interactions influence the electron wavefunction of the atom. This finally results again in atomic observables.

If we consider cosmic fields that couple directly to nucleons, we obtain exactly the same potentials [Eqs.~(\ref{eq: all potentials}--\ref{eq: all potentials Th})] as when we treated the coupling to electrons. The only difference is that now the operators $\bs{\sigma}_n$, and $\bs{p}_n$ act on the nucleon- instead of the electron-wavefunctions. Additionally, the relativistic suppression for nucleons are even stronger ($m_p,\,m_n\gg m_e$). The induced nuclear moments can be derived through exactly the same procedure as before.

The induced nuclear magnetic dipole $\langle\tilde{\bs{\mu}}\rangle$ and electric quadrupole moment $\langle \tilde{Q}_{ij}\rangle$ lead to an anisotropic (time-varying) component to the usual atomic hyperfine coupling. This anisotropy can for example be measured through sidereal variations of the hyperfine transition frequency \cite{nanda2025demonstration,humphrey2003testing} or by searching for a lifting of rotational symmetry at zero field \cite{nowak2024cpt}. Since all the interaction potentials contributing to $\langle\tilde{\bs{\mu}}\rangle$ and $\langle \tilde{Q}_{ij}\rangle$ depend explicitly on the total nucleon spin $\bs{\sigma}_n$, only unpaired valence nucleons can contribute to the interaction to leading order.

The nuclear electric dipole moment $\langle\tilde{\bs{d}}\rangle$ is usually difficult to observe. The electron wavefunction shields it from any static electric fields through the Schiff theorem \cite{schiff1963measurability, sandars1965electric}. If, however, through the interaction with cosmic fields, $\langle\tilde{\bs{d}}\rangle$ oscillates reasonably fast, this shielding is weakened \cite{flambaum2023screening}. Otherwise, one can also utilize the mean square radius of the nuclear electric dipole moment (also known as the Schiff moment) \cite{ginges2004violations}. It as well circumvents some of the shielding. The nuclear Schiff moment $\langle \tilde{\bs{S}}\rangle$ has the same transformation properties as the electric dipole moment (see Table~\ref{tab: P T k}) and is therefore sensitive to exactly the same terms. The atomic interaction potential of the coupling between the electron and nuclear moments is given by: $V_{eN}=e4\pi\big[\langle \tilde{\bs{d}} \rangle+\langle\tilde{\bs{S}}\rangle
\big]\cdot\bs{\nabla}\delta(\bs{r})$, where $\bs{r}$ is the position operator of the electron.~\cite{ginges2004violations}. This interaction is $P$-odd $T$-odd and as such induces atomic dipole moments. The probability of the electron to be close to the nucleus $\delta(\bs{r})$ and the momentum $-i\bs{\nabla}$ both scale approximately proportional with the atomic number $Z$ \cite{khriplovich1991parity}. One can also expect additional scalings from $\langle\tilde{\bs{d}}\rangle$, and $\langle\tilde{ \bs{S}}\rangle$ themselves. It is evidently that the sensitivity to this subset of nuclear cosmic field interactions scales strongly with the atomic number $Z$ making heavy atoms the preferred system. Because $V_{eN}$ induces an atomic electric dipole moment, searches for oscillating atomic EDMs are at the same time sensitive to the interactions discussed here and the ones from Sec.~\ref{sec: The induced electric dipole moment}. Rydberg atoms, on the other hand, are mostly insensitive to nuclear moments.

\subsection{The Nuclear Anapole Mmoment} \label{sec: The nuclear anapole moment}
The nuclear anapole moment $\bs{\mathrm{a}}$ is a toroidal moment that is usually discussed in the context of atomic parity violation \cite{guena2005atomic}. As is the case for the electric dipole moment, all $P$-odd interaction potentials contribute to the anapole moment. However, due to the different time-reversal properties of $\bs{d}$ and $\bs{\mathrm{a}}$, the exact dependency on the terms differs. From the considerations in Appendix \ref{app: Derivation of the...} follows:
\begin{widetext}
\begingroup%this makes sure that the equation doesnt jump to the next page
\setlength{\abovedisplayskip}{0pt}
\setlength{\belowdisplayskip}{0pt}
\begin{align}
  &\langle\tilde{\bs{\mathrm{a}}}\rangle_\phi=g_\phi\Big[-{\alpha'}^s_{ \mathrm{a} \phi_2}(\bs{\nabla}\dot{\phi})+{\alpha}^v_{ \mathrm{a} \phi_2}\big[(\bs{\nabla}{\phi})\times\langle\bs{I}\rangle\big]-{\alpha'}^t_{ \mathrm{a} \phi_2}\tilde{\mathcal{Q}}\cdot(\bs{\nabla}\dot{\phi})\Big]\\[8pt]
 & \langle \tilde{\bs{ \mathrm{a} }}\rangle_a=g_a {\alpha}_{ \mathrm{a} a_3}\langle\bs{I}\rangle\dot{a}\\[8pt]
  &\langle \tilde{\bs{ \mathrm{a}} }\rangle_{A'}=g_{A'}\bigg[{\alpha}^s_{ \mathrm{a} A'_2}{\bs{A}}'-\Big[{\alpha'}^s_{ \mathrm{a} A'_6}+{\alpha'}^s_{ \mathrm{a} A'_7}\Big]\ddot{\bs{A}}'-{\alpha'}^s_{ \mathrm{a} A'_4}(\bs{\nabla}\dot{A}'_0)\\
 &\hspace{33pt}+\Big[-{\alpha'}^v_{ \mathrm{a} A'_2}+{\alpha}^v_{ \mathrm{a} A'_6}+{\alpha}^v_{ \mathrm{a} A'_7}\Big]\big[\dot{\bs{A}}'\times\langle\bs{I}\rangle\big]+{\alpha}^v_{ \mathrm{a} A'_4}\big[(\bs{\nabla}A'_0)\times\langle\bs{I}\rangle\big]\nonumber\\
   &\hspace{33pt}+{\alpha}^t_{ \mathrm{a} A'_2}\tilde{\mathcal{Q}}\cdot{\bs{A}}'-\Big[{\alpha'}^t_{ \mathrm{a} A'_6}+{\alpha'}^t_{ \mathrm{a} A'_7}\Big]\tilde{\mathcal{Q}}\cdot\ddot{\bs{A}}' -{\alpha'}^t_{ \mathrm{a} A'_4}\tilde{\mathcal{Q}}\cdot(\bs{\nabla}\dot{A}'_0)\bigg]\nonumber\\[8pt]
  &\langle \tilde{\bs{ \mathrm{a}} }\rangle_{Z'}=g_{Z'}\bigg[
   { \alpha}_{ \mathrm{a} Z'_2}\langle\bs{I}\rangle{Z}'_0-{\alpha'}_{ \mathrm{a} Z'_4}\langle\bs{I}\rangle\ddot{Z}'_0-{\alpha'}_{ \mathrm{a} Z'_6}\langle\bs{I}\rangle(\bs{\nabla}\cdot\dot{\bs{Z}}'_0)+{\alpha}^s_{ \mathrm{a} Z'_7}(\bs{\nabla}\times{\bs{Z}}')\\
   &\hspace{33pt}-{\alpha'}^v_{ \mathrm{a} Z'_7}\big[(\bs{\nabla}\times\dot{\bs{Z}}')\times\langle\bs{I}\rangle\big]+{\alpha}^t_{ \mathrm{a} Z'_7}\tilde{\mathcal{Q}}\cdot(\bs{\nabla}\times{\bs{Z}}')\bigg]\nonumber\\[8pt]
   &\langle \tilde{\bs{ \mathrm{a} }}\rangle_{\Theta}=g_{\Theta}\bigg[\Big[-{\alpha'}^s_{ \mathrm{a} \Theta_3}+{\alpha}^s_{ \mathrm{a} \Theta_5}+{\alpha}^s_{ \mathrm{a} \Theta_6}\Big]\dot{\bs{\theta}}^E-{\alpha'}_{\Theta_7}\langle\bs{I}\rangle(\bs{\nabla}\cdot\dot{\bs{\theta}}^B)
         +{\alpha}^s_{ \mathrm{a} \Theta_8}(\bs{\nabla}\times{\bs{\theta}}^E)\\
   &\hspace{33pt}+{\alpha}^v_{ \mathrm{a} \Theta_3}\big[{\bs{\theta}}^E\times\langle\bs{I}\rangle\big]-\Big[{\alpha'}^v_{ \mathrm{a} \Theta_5}+{\alpha'}^v_{ \mathrm{a} \Theta_6}\Big]\big[\ddot{\bs{\theta}}^E\times\langle\bs{I}\rangle\big]-{\alpha'}^v_{ \mathrm{a} \Theta_8}\big[(\bs{\nabla}\times \dot{\bs{\theta}}^E)\times\langle\bs{I}\rangle\big]\nonumber\\
   &\hspace{33pt}+\Big[-{\alpha'}^t_{ \mathrm{a} \Theta_3}+{\alpha}^t_{ \mathrm{a} \Theta_5}+{\alpha}^t_{ \mathrm{a} \Theta_6}\Big]\tilde{\mathcal{Q}}\cdot\dot{\bs{\theta}}^E+{\alpha}^t_{ \mathrm{a} \Theta_8}\tilde{\mathcal{Q}}\cdot(\bs{\nabla}\times{\bs{\theta}}^E)\bigg]\nonumber
\end{align} 
\endgroup
\end{widetext}
Here, $\bs{I}$ is the nuclear spin, $\tilde{\mathcal{Q}}$ is the nuclear spin quadrupole, and $\langle...\rangle$ refers to the expectation value in the nucleon wavefunctions.

We can see that the same operators contribute to $\langle\tilde{\bs{\mathrm{a}}}\rangle$ as to $\langle\tilde{\bs{d}}\rangle$ and $\langle\tilde{\bs{S}}\rangle$, but with the opposite frequency scaling ($\alpha\leftrightarrow\alpha'$). This is especially relevant because of the large energy difference $\omega_{nk}$ between states of different parities in nuclei, the polarizabilities $\alpha'$ are strongly suppressed compared to polarizabilities $\alpha$.

One other peculiarity of the nuclear anapole moment is that it contains with $ \alpha_{\mathrm{a}Z'_2}\langle\bs{I}\rangle{Z}'_0$ a term, where the cosmic field is neither polarized, nor does it involve spatial or time derivatives of it. This means that a static cosmic field would modify the value of the intrinsic anapole moment compared to its expectation value in the standard model. 

The nuclear anapole moment interacts with the atomic electrons through the potential  $V_{eN}=\bs{\mathrm{a}}\cdot\bs{\alpha}\delta(\bs{r})$. Like the interaction with the electric dipole moment and Schiff moment discussed above, $\bs{\alpha}\delta{(\bs{r})}$ scales with $Z^2$. Anapole moments can be measured by driving highly forbidden transitions \cite{wood1997measurement}.

\section{Conclusion}
We have shown how cosmic fields can induce atomic electric dipole $\bs{d}$, magnetic dipole $\bs{\mu}$, electric quadrupole $\bs{Q}$, as well as nuclear Schiff $\tilde{\bs{S}}$ and anapole $\tilde{\bs{\mathrm{a}}}$ moments. For this, we considered fields that couple to the electron or nucleons through either scalar, $\phi$ pseudoscalar $a$, vector $A'$, axial vector $Z'$, or tensor $\Theta$ couplings. We found that each atomic observable possesses some sensitivity to every type of cosmic field coupling. Similar to electromagnetic fields, cosmic fields can polarize and magnetize atomic systems. 

We discussed which aspects need to be considered to evaluate the relative scaling of the different terms. It is impossible to make a general statement about which interaction will be the most promising, as this depends on many different properties of the system. Firstly, which types of couplings are of relevance depends on the type of new physics theory one wants to test. Further, depending on the production mechanism and evolution of the cosmic field, terms that depend on its polarization might be strongly suppressed. At the same time, the properties of the atomic system are of great importance, such as its angular momentum and transition elements. Some terms are relativistically suppressed in light atoms, but might become relevant in heavy ones. Finally, situations can arise where the energy difference between atomic levels is in resonance with the mass of the new boson, which leads to large enhancements.

We hope this work can serve as an overview of the different ways cosmic fields can interact with atomic systems. This can be a guide to reevaluate existing measurements and develop new ones with optimized sensitivities to couplings that have received less attention so far. 

\section{Acknowledgments}
This research was financed in whole or in part by Agence Nationale de la Recherche (ANR) under the projects ANR-21-CE30 -0028-0 and ANR-24-CE96-0002. B.M.R. was supported by the Australian Research Council (ARC) DECRA Fellowship DE210101026, and the UQ Fellowship of the Big Questions Institute.
FK acknowledges funding by the Deutsche Forschungsgemeinschaft (DFG, German Research Foundation) under Germany’s Excellence Strategy – EXC-2123 QuantumFrontiers – 390837967.

A CC-BY public copyright license has been applied by the authors to the present document and will be applied to all subsequent versions up to the Author Accepted Manuscript arising from this submission, in accordance with the grant's open access conditions.

\appendix
\begin{widetext}
\section{Derivation of the $P$, $T$-Symmetry Dependencies of scalar, vector, and tensor interactions} \label{app: Derivation of the...}
In the following, we derive to which field interactions electric and magnetic dipole moments and the electric quadrupole moment are sensitive to. The procedure closely resembles the one presented in Ref.~\cite{lahs2024polarizabilities}. 
\subsection{Scalar Interactions}
If there is some external field that atoms or molecules can couple to through a semiclassical scalar interaction, this can be expressed as a potential $V^\varphi= \big[\mathcal{I}+i\mathcal{K}\big]\varphi(t)$. Here, $ {\mathcal{I}}$ and $\mathcal{K}$ are Hermitian scalar atomic operators. $\varphi(t)$ is a (time-dependent) scalar that quantifies the strength of the external field. $\varphi$ should not be confused with $\phi$ from above. $\varphi$ represents any quantity $f(\Xi)$ of rank 0 (See Table \ref{tab:catalog}). The time dependency of $\varphi$ can be expressed as: \\
\begin{align*}
 {\varphi}(t)&= \mathrm{Re}\Big[ c e^{-i\omega_\varphi t}\Big]
 \end{align*}
This form also allows for the treatment of static fields by setting $\omega_\varphi=0$. 
We can now perform a perturbative treatment to calculate the induced electric dipole moment $\langle \bs{d} \rangle $. Because the perturbation potential $V^\varphi$ is a sum of potentials periodic in time (see \cite{langhoff1972aspects,landau2013quantum6}), the induced dipole moment is given by:
\begin{align}
\langle d_i\rangle =-\sum_{k}&\left[\frac{\br{n}d_i\ke{k} \br{k} {\mathcal{I}} c\ke{n}}{\omega_{k n}-\omega_\varphi}+\frac{\br{n} {\mathcal{I}} c\ke{k}\br{k}d_i\ke{n} }{\omega_{k n}+\omega_\varphi}\right] e^{-i \omega_\varphi t}\nonumber\\
+&\left[\frac{\br{n}d_i\ke{k} \br{k}\mathcal{I} c^*\ke{n}}{\omega_{k n}+\omega_\varphi}+\frac{\br{n}{\mathcal{I}} c^* \ke{k}\br{k}d_i\ke{n} }{\omega_{k n}-\omega_\varphi}\right] e^{i \omega_\varphi t}\nonumber\\
+&\left[\frac{\br{n}d_i\ke{k} \br{k} {i\mathcal{K}} c\ke{n}}{\omega_{k n}-\omega_\varphi}-\frac{\br{n} {i\mathcal{K}} c\ke{k}\br{k}d_i\ke{n} }{\omega_{k n}+\omega_\varphi}\right] e^{-i \omega_\varphi t}\nonumber\\
+&\left[\frac{\br{n}d_i\ke{k} \br{k}i\mathcal{K} c^*\ke{n}}{\omega_{k n}+\omega_\varphi}-\frac{\br{n}{i\mathcal{K}} c^* \ke{k}\br{k}d_i\ke{n} }{\omega_{k n}-\omega_\varphi}\right] e^{i \omega_\varphi t}\nonumber
\end{align}
Here, $\ke{n}$, $\ke{k}$ are atomic states and $\omega_{kn}$ is the energy difference between them. $\langle d_i\rangle$ describes the $i$th spatial component of the vector $\langle \bs{d} \rangle $. The expression above is only correct far from resonance $\left(|\omega_\varphi|\approx|\omega_{kn}|\right)$. Close to resonance, one needs to take the natural linewidth of the states into account \cite{vexiau2017dynamic, lahs2024polarizabilities}.
By expanding the fractions and identifying $-i\omega_\varphi c e^{-i \omega_\varphi t}+i\omega_\varphi c^* e^{i \omega_\varphi t}$ with the time derivative of $\varphi$, we can rewrite this expression as:
\begin{align*}
 \langle d_i \rangle = &-2\mathrm{Re}\Bigg[\sum_k\frac{\br{n}d_i\ke{k}\br{k}{\mathcal{I}} \ke{n}}{\omega_{k n}^2-\omega_\varphi^2}\Big[\omega_{kn}\varphi+i\dot{\varphi}\Big]\Bigg]\\
 &+2\mathrm{Im}\Bigg[\sum_k\frac{\br{n}d_i\ke{k}\br{k}{\mathcal{K}} \ke{n}}{\omega_{k n}^2-\omega_\varphi^2}\Big[\omega_{kn}\varphi+i\dot{\varphi}\Big]\Bigg]\\
 &=\mathrm{Re} 
 \Big[{\beta}^{d\mathcal{I}}_{i}\Big]\varphi-\mathrm{Im}\Big[{\beta'}^{d\mathcal{I}}_{i}\Big]\dot{\varphi}+\mathrm{Im} 
 \Big[{\beta}^{d\mathcal{K}}_{i}\Big]\varphi-\mathrm{Re}\Big[{{\beta}'}^{d\mathcal{K}}_{i}\Big]\dot{\varphi}
\end{align*}
$\mathrm{Re}(.)$ is the real and $\mathrm{Im}(.)$ the imaginary part. Above, we introduced the (generalized) polarizabilities:
\begin{align*}
{\beta}^{d\mathcal{I}}_{i}&=-2\sum_k\frac{\br{n}d_i\ke{k}\br{k}\mathcal{I}\ke{n}}{\omega_{kn}^2-\omega_\varphi^2}\omega_{kn},\quad{{\beta}}^{d\mathcal{K}}_{i}=-2\sum_k\frac{\br{n}d_i\ke{k}\br{k}\mathcal{K}\ke{n}}{\omega_{kn}^2-\omega_\varphi^2}\omega_{kn} \\
{\beta'}^{d\mathcal{I}}_{i}&=-2\sum_k\frac{\br{n}d_i\ke{k}\br{k}\mathcal{I}\ke{n}}{\omega_{kn}^2-\omega_\varphi^2},\;\,\quad{{\beta}'}^{d\mathcal{K}}_{i}=-2\sum_k\frac{\br{n}d_i\ke{k}\br{k}\mathcal{K}\ke{n}}{\omega_{kn}^2-\omega_\varphi^2} \nonumber
\end{align*}
%\FK{I think I'd find it easier to follow without the introduction of these objects. Can you directly move on to the $\alpha$'s, which you also use in the main text?}\SL{I prefer it like this. I think it makes it easier to follow the structure of the equations. The case for vector interactions later is more complicated. Doing the scalar process like this step by step, makes it easier to understand the vector polarizabilities later on.}
Likewise, we can perform the same derivation for the magnetic dipole and electric quadrupole moment, which leads to:
\begin{align*}
\langle \mu_i\rangle &= \br{n}\mu_i\ke{n} +\mathrm{Re} 
 \Big[{\beta}^{\mu \mathcal{I}}_{i}\Big]\varphi-\mathrm{Im}\Big[{\beta'}^{\mu\mathcal{I}}_{i}\Big]\dot{\varphi}+\mathrm{Im} 
 \Big[{\beta}^{\mu \mathcal{K}}_{i}\Big]\varphi-\mathrm{Re}\Big[{\beta'}^{\mu\mathcal{K}}_{i}\Big]\dot{\varphi}\\
 \langle Q_{ij} \rangle &= \br{n}Q_{ij}\ke{n} +\mathrm{Re} 
 \Big[{\beta}_{ij}^{Q\mathcal{I}}\Big]\varphi-\mathrm{Im}\Big[{\beta'}^{Q\mathcal{I}}_{ij}\Big]\dot{\varphi}+\mathrm{Im} 
 \Big[{\beta}^{Q\mathcal{K}}_{ij}\Big]\varphi-\mathrm{Re}\Big[{\beta'}^{Q\mathcal{K}}_{ij}\Big]\dot{\varphi}
\end{align*}
Different from the electric dipole moment, the magnetic dipole and electric quadrupole possesses zeroth order contributions that are independent of the external fields.

\subsection{Analyzing Interactions According to their Transformations under $P$ and $T$}

\begin{table}[]
\centering
\large
\begin{tabular}{c|cccc|ccccc}
  & $\bs{\sigma}$& $\bs{p}$& $\bs{J}$& $\mathcal{Q}_{ij}$& $\bs{d}$ & $\bs{\mu}$ & $Q_{ij}$& $\bs{S}$ & $\bs{\mathrm{a}}$ \\ \hline
$P$ & +                      & -            & +            & +                  & -    & +     & +    & -    & -         \\
$T$ & -                      & -            & -            & +                  & +    & -     & +    & +    & -         \\
$k$ & 1                      & 1            & 1            & 2                  & 1    & 1     & 2    & 1    & 1         
\end{tabular}\caption{The transformation behaviour under parity $P$, time reversal $T$, and the rank $k$ of the different relevant atomic operators and multipole moments is stated.}\label{tab: P T k}
\end{table}

According to the Wigner-Eckart theorem, the expectation value of a vector operator needs to be necessarily proportional to $\langle\bs{J}\rangle$, the total angular momentum in the system, while the expectation value of a symmetric traceless tensor like $Q_{ij}$ is necessarily proportional to the angular momentum quadrupole  $
\mathcal{Q}_{ij}\equiv\br{n}\tfrac{1}{2}\Big[J_i J_j+J_j J_i-\tfrac{2}{3}\delta_{ij}\sum_lJ_l J_l\Big]\ke{n}$ \cite{varshalovich1988quantum}. Under these considerations, one can write the polarizability tensors $\beta$ in respect to real constants $\alpha$:
\begin{align*}
    \mathrm{Re}\Big[{\beta}^{d\mathcal{I}}_{i}\Big]={\alpha}_{d\mathcal{I}}\langle J_i\rangle , \quad \mathrm{Im}\Big[{\beta'}^{d\mathcal{I}}_{i}\Big]={{\alpha}'}_{d\mathcal{I}}\langle J_i\rangle,\quad \mathrm{Im}\Big[{\beta}^{d\mathcal{K}}_{i}\Big]={\alpha}_{d\mathcal{K}}\langle J_i\rangle ,\quad \mathrm{Re}\Big[{\beta'}^{d\mathcal{K}}_{i}\Big]={{\alpha}'}_{d\mathcal{K}}\langle J_i\rangle
\end{align*}
The constants for $\bs{\mu}$ and $Q_{ij}$ follow analogously, with the difference that for $Q_{ij}$, the vector $\langle J_i\rangle$ needs to be replaced by the tensor $\mathcal{Q}_{ij}$.

In \cite{lahs2024polarizabilities}, we demonstrated how polarizability tensors can vanish if their transformations under parity $P$ and time reversal $T$ are considered. Terms will only be non-vanishing if the product of the involved multipole ($d,\,\mu,\,Q$) and operator ($\mathcal{I},\, \mathcal{K}$) is $P$-even. Further, the following identity, which follows from the Wigner-Eckart theorem, is fulfilled for the complex conjugate of the expectation value of any spherical tensor operator $X_q^k$ \cite{stephens1993time}:
\begin{align}
  \langle X_q^k\rangle^*=(-1)^{k+\tau}\,\langle X_q^k\rangle \label{eq: Qqk}
\end{align}
Here $k$ is the rank of the tensor operator and $\tau\in\{0,1\}$ describes if it is even or odd under time reversal. For $\beta_i$ that has a rank of $k=1$, this implies that if the product of the involved dipole and atomic operator is $T$-even, then the resulting $\beta$ tensor will be imaginary. If the product is $T$-odd, $\beta$ will be real.

In this context, it is convenient to decompose the atomic operators $\mathcal{I}$ and $\mathcal{K}$ into sub-components:
\begin{align*}
  \mathcal{I}&=I+I^P+I^T+I^{PT}\\
  \mathcal{K}&=K+K^P+K^T+K^{PT}
\end{align*}
Here, the upper indices signal under which symmetries the component is odd. The symmetries of the multipole operators are indicated in Table~\ref{tab: P T k}.
Applying the rules to all theoretically possible operators gives us the following nonvanishing contributions:
\begin{align*}
\langle \bs{d}\rangle&=\Big[{\alpha}_{d I^{PT}}+{\alpha}_{d K^{P}}\Big]\langle\bs{J}\rangle\varphi-\Big[{\alpha'}_{d I^{P}}+{\alpha'}_{d K^{PT}}\Big]\langle\bs{J}\rangle\dot{\varphi}\\
\langle \bs{\mu}\rangle&={\alpha}_{\mu}\langle\bs{J}\rangle+\Big[{\alpha}_{\mu I^{}}+{\alpha}_{\mu K^{T}}\Big]\langle\bs{J}\rangle\varphi-\Big[{\alpha'}_{\mu I^{T}}+{\alpha'}_{\mu K^{}}\Big]\langle\bs{J}\rangle\dot{\varphi}\\
\langle Q_{ij}\rangle&={\alpha}_{Q}{\mathcal{Q}}+\Big[{\alpha}_{Q I^{}}+{\alpha}_{d K^{T}}\Big]{\mathcal{Q}_{ij}}\varphi-\Big[{\alpha'}_{d I^{T}}+{\alpha'}_{d K^{}}\Big]
{\mathcal{Q}_{ij}}\dot{\varphi}
\end{align*}
where $\alpha_\mu \langle\bs{J}\rangle=\br{n}\bs{\mu}\ke{n}$ and $\alpha_Q {\mathcal{Q}_{ij}}=\br{n}Q_{ij}\ke{n}$. The term $\alpha'_{\mu I^T}$ for example is defined as $\alpha'_{\mu I^T}=-2\mathrm{Im}\sum_k\frac{\br{n}\mu_i\ke{k}\br{k}I^T\ke{n}}{\omega_{kn}^2-\omega_\varphi^2}$.
\subsection{Vector Interaction}\label{sec: Vector interaction}
Now, instead of a scalar, we assume there to be a vector interaction with an external field. We express this as
$V^\mathcal{V}=\big[ \bs{\mathcal{I}}+i\bs{\mathcal{K}}\big] \cdot \bs{\mV}(t)$. $ \bs{\mathcal{I}}$ and $\bs{\mathcal{K}}$ are Hermitian (3 component) vector operators and $\bs{\mV}(t)$ is a (classical) vector field. We write it as:
\begin{align*}
 \bs{\mV}(t)&= \mathrm{Re} \Big[\bs{\nu} e^{-i\omega_\mV t}\Big]
 \end{align*}
$\bs{\nu}$ is a complex vector that defines the magnitude and polarization of the field. The derivation of the induced multipole moments follows now quite analogously to before. From it, we obtain:
\begin{align*}
 \langle d_i \rangle =& -2\mathrm{Re}\sum_k\frac{\br{n}d_i\ke{k}\br{k}{\mathcal{I}}_j \ke{n}}{\omega_{k n}^2-\omega_\mV^2}\big[\omega_{kn}\mV_j+i\dot{\mV}_j\big]\\
 &+2\mathrm{Im}\sum_k\frac{\br{n}d_i\ke{k}\br{k}{\mathcal{K}}_j \ke{n}}{\omega_{k n}^2-\omega_\mV^2}\big[\omega_{kn}\mV_j+i\dot{\mV}_j\big]\\
 &=\mathrm{Re} 
 \Big[{\beta}^{d\mathcal{I}}_{ij}\Big]\mV_j-\mathrm{Im}\Big[{\beta'}^{d\mathcal{I}}_{ij}\Big]\dot{\mV}_j+\mathrm{Im} 
 \Big[{\beta}^{d\mathcal{K}}_{ij}\Big]\mV_j-\mathrm{Re}\Big[{\beta'}^{d\mathcal{K}}_{ij}\Big]\dot{\mV}_j\\
 \text{and}\quad&\\
 \langle \mu_i\rangle &= \br{n}\mu_i\ke{n} +\mathrm{Re} 
 \Big[{\beta}^{\mu\mathcal{I}}_{ij}\Big]\mV_j-\mathrm{Im}\Big[{\beta'}^{\mu\mathcal{I}}_{ij}\Big]\dot{\mV}_j+\mathrm{Im}
 \Big[{\beta}^{\mu\mathcal{K}}_{ij}\Big]\mV_j-\mathrm{Re}\Big[{\beta'}^{\mu\mathcal{K}}_{ij}\Big]\dot{\mV}_j\\
 \langle Q_{ij} \rangle &= \br{n}Q_{ij}\ke{n} +\mathrm{Re} 
 \Big[{\beta}^{Q\mathcal{I}}_{ij l }\Big]\mV_ l -\mathrm{Im}\Big[{\beta'}^{Q\mathcal{I}}_{ij l }\Big]\dot{\mV}_ l \\
 &\hspace{53pt}+ \mathrm{Im}
 \Big[{\beta}^{Q\mathcal{K}}_{ij l }\Big]\mV_ l -\mathrm{Re}\Big[{\beta'}^{Q\mathcal{K}}_{ij l }\Big]\dot{\mV}_ l 
\end{align*}
Analogous to the atomic operators in the scalar interaction, we can express $\bs{\mathcal{I}}$ and $\bs{\mathcal{K}}$ as a sum of vectors transforming differently under $P$ and $T$:
\begin{align*}
  \bs{\mathcal{I}}&={\bs{I}}+{\bs{I}}^{P}+{\bs{I}}^{T}+{\bs{I}}^{PT}\\
  \bs{\mathcal{K}}&={\bs{K}}+{\bs{K}}^{P}+{\bs{K}}^{T}+{\bs{K}}^{PT}
\end{align*}
As before, we have the requirement that the product of the multipole and the atomic operator has to be $P$-even. The polarizabilities $\beta_{ij}$ are general rank-2 tensors. As such, these can be decomposed into a sum of scalar, a vector, and a symmetric tensor component. The form of this decomposition depends on whether the product of the involved multipole and atomic operator $\bs{\mathcal{I}}$ or $\bs{\mathcal{K}}$ is $T$-even, or $T$-odd (see eq.~\ref{eq: Qqk}):
\begin{align*}
 T\text{-even:  }\prescript{}{}{\beta}_{ij}&=\delta_{ij}\; \alpha^{s}+i\varepsilon_{ij l }\mathcal{J}_ l \; \alpha^{v}+\mathcal{Q}_{ij}\; \alpha^{t}\\
 T\text{-odd:  }\prescript{}{}{\beta}_{ij}&=i\delta_{ij}\; \alpha^{s}+\varepsilon_{ij l }\mathcal{J}_ l \; \alpha^{v}+i\mathcal{Q}_{ij}\; \alpha^{t}
\end{align*}
Again, all $\alpha$ are real constants. $\varepsilon_{ij l }$ is the Levi-Civita symbol. The same relations apply for $\beta'$. 

${\beta}_{ij l }$ is a rank-3 tensor that is symmetric in its first two indices. It can be decomposed into\cite{itin2022decomposition}:
\begin{align*}
 T\text{-even:  }{\beta}_{ij l }&=i\mathcal{O}_{ l  ij}\,\alpha^{t_s, \ell}+i\mathcal{M}_{ l  ij}\,\alpha^{v_m, \ell}+\mathcal{W}_{ l  ij}\,\alpha^{t_m, \ell}\\
 T\text{-odd:  }{\beta}_{ij l }&=\mathcal{O}_{ l  ij}\,\alpha^{t_s, \ell}+\mathcal{M}_{ l  ij}\,\alpha^{v_m, \ell}+i\mathcal{W}_{ l  ij}\,\alpha^{t_m, \ell}
\end{align*}
The three tensors contained in this decomposition are defined as follows:
\begin{align*}
\mathcal{T}_{ij l }&=\tfrac{1}{6}(J_iJ_jJ_ l  +J_jJ_ l  J_i +J_ l  J_iJ_j+J_jJ_iJ_ l +J_ l  J_jJ_i+J_iJ_ l  J_j)\nonumber\\
\mathcal{O}_{ij l }&=\langle\mathcal{T}_{ij l }-\tfrac{1}{5}(\mathcal{T}_{inm}\delta_{nm}\delta_{j l }+\mathcal{T}_{njm}\delta_{nm}\delta_{i l }+\mathcal{T}_{nm l }\delta_{nm}\delta_{ij})\rangle\nonumber \\
\mathcal{M}_{ij l }&=\tfrac{1}{4}\langle2\delta_{j l }J_i -\delta_{ij}J_ l -\delta_{i l }J_j\rangle\nonumber\\
\mathcal{W}_{ij l }&=\tfrac{1}{2}(\delta_{jl}\varepsilon_{im l }+\delta_{ l  l}\varepsilon_{imj})\mathcal{Q}_{lm}\nonumber
\end{align*}
Above's considerations result in the following induced moments:
\begin{align*}
   \langle \bs{d} \rangle = &\Big[{\alpha}_{d I^{P}}^{s}+{\alpha}_{d K^{PT}}^{s}\Big]\bs{\mV}-\Big[{\alpha'}_{d I^{P}}^{v}+{\alpha'}_{d K^{PT}}^{v}\Big] (\dot{\bs{\mV}}\times\langle\bs{J}\rangle)+\Big[{\alpha}_{d I^{P}}^{t}+{\alpha}_{d K^{PT}}^{t}\Big]{\mathcal{Q}}\cdot\bs{\mV}\\
   -&\Big[{\alpha'}_{d I^{PT}}^{s}+{\alpha'}_{d K^{P}}^{s}\Big]\dot{\bs{\mV}}+\Big[{\alpha}_{d I^{PT}}^{v}+{\alpha}_{d K^{P}}^{v}\Big] (\bs{\mV}\times\langle\bs{J}\rangle)-\Big[{\alpha'}_{d I^{PT}}^{t}+{\alpha'}_{d K^{P}}^{t}\Big]{\mathcal{Q}}\cdot \dot{\bs{\mV}}\\
   \langle \bs{\mu} \rangle = &{\alpha}_{\mu}^{}\langle\bs{J}\rangle+\Big[{\alpha}_{\mu I^{T}}^{s}+{\alpha}_{\mu K^{}}^{s}\Big] \bs{\mV}-\Big[{\alpha'}_{\mu I^{T}}^{v}+{\alpha'}_{\mu K^{}}^{v}\Big] (\dot{\bs{\mV}}\times\langle\bs{J}\rangle)+\Big[{\alpha}_{\mu I^{T}}^{t}+{\alpha}_{\mu K^{}}^{t}\Big] \mathcal{Q}\cdot\bs{\mV}\nonumber\\
   -&\Big[{\alpha'}_{\mu I^{}}^{s}+{\alpha'}_{\mu K^{T}}^{s}\Big] \dot{\bs{\mV}}+\Big[{\alpha}_{\mu I^{}}^{v}+{\alpha}_{\mu K^{T}}^{v}\Big] (\bs{\mV}\times\langle\bs{J}\rangle)-\Big[{\alpha'}_{\mu I^{}}^{t}+{\alpha'}_{\mu K^{T}}^{t}\Big] \mathcal{Q}\cdot \dot{\bs{\mV}}\\
  \langle Q_{ij}\rangle =&\,{\alpha}_{Q}^{}\mathcal{Q}_{ij}- \left(\Big[{\alpha'}_{Q I^{}}^{t_s}+{\alpha'}_{Q K^{T}}^{t_s}\Big]\mathcal{O}_{ l  ij}+\Big[{\alpha'}_{Q I^{}}^{v_m}+{\alpha'}_{Q K^{T}}^{v_m}\Big]\mathcal{M}_{ l  ij}\right) \dot{\mV}_ l \nonumber\\
 +&\left(\Big[{\alpha}_{Q I^{T}}^{t_s}+{\alpha}_{Q K^{}}^{t_s}\Big]\mathcal{O}_{ l  ij}+\Big[{\alpha}_{Q I^{T}}^{v_m}+{\alpha}_{Q K^{}}^{v_m}\Big]\mathcal{M}_{ l  ij}\right) {\mV}_ l \nonumber\\
 +&\Big[{\alpha}_{Q I^{}}^{t_m}+{\alpha}_{Q K^{T}}^{t_m}\Big]\mathcal{W}_{ l  ij} {\mV}_ l -\Big[{\alpha'}_{Q I^{T}}^{t_m}+{\alpha'}_{Q K^{}}^{t_m}\Big]\mathcal{W}_{ l  ij} \dot{\mV}_ l 
\end{align*}

\subsection{Energy Shifts} \label{sec: energy shifts}
The direct energy shift through the cosmic field is simply given by the expectation value:
\begin{align*}
  \Delta E=&\mathrm{Re}\Big[\br{n}\big([\mathcal{I}+i\mathcal{K}]\varphi(t)+[ \bs{\mathcal{I}}+i\bs{\mathcal{K}}] \cdot \bs{\mV}(t)\big)\ke{n}\Big]\\
\end{align*}
Only $P$-even operators can contribute inside the states $\ke{n}$, which have a definite parity. To find the conditions on $T$, we need to apply Eq.~(\ref{eq: Qqk}). This yields:
\begin{align*}
  \Delta E=&\langle I\rangle\varphi(t)+\langle K^T\rangle \varphi(t)+\langle\bs{I}^T\rangle \cdot \bs{\mV}(t)+\langle\bs{K}\rangle \cdot \bs{\mV}(t)
\end{align*}
\subsection{Fields with more Complicated Time-Dependencies}
The derivation above can only describe cosmic fields that are either static or undergo harmonic oscillations. If one wants to describe cosmic fields with more complicated time-dependencies, the derivation above can be slightly modified. This is, for example, relevant for UBDM fields that additionally undergo sidereal variations. To account for multiple time-dependencies, an arbitrary field can be developed as a Fourier series:
\begin{align*}
   {\varphi}(t)&=\sum_{\ell}\varphi^\ell=\sum_{\ell} \mathrm{Re}\Big[ c^\ell e^{-i\omega_\varphi^\ell t}\Big]\\
   \bs{\mV}(t)&=\sum_{\ell}\bs{\mV}^\ell=\sum_{\ell} \mathrm{Re} \Big[\bs{\nu}^\ell e^{-i\omega_\mV^\ell t}\Big]
\end{align*}
In the order of perturbation we discuss here, there will be no interference between different terms in the sum. One can therefore simply sum the different components. For example, the induced electric dipole moment of an arbitrary vector field can be written as:
\begin{align*}
   \langle \bs{d} \rangle =&\Big[{\alpha}_{d I^{P}}^{s,\ell}+{\alpha}_{d K^{PT}}^{s,\ell}\Big]\bs{\mV}^\ell-\Big[{\alpha'}_{d I^{P}}^{v,\ell}+{\alpha'}_{d K^{PT}}^{v,\ell}\Big] (\dot{\bs{\mV}}^\ell\times\langle\bs{J}\rangle)+\Big[{\alpha}_{d I^{P}}^{t,\ell}+{\alpha}_{d K^{PT}}^{t,\ell}\Big]{\mathcal{Q}}\cdot\bs{\mV}^\ell\\
   -&\Big[{\alpha'}_{d I^{PT}}^{s,\ell}+{\alpha'}_{d K^{P}}^{s,\ell}\Big]\dot{\bs{\mV}}^\ell+\Big[{\alpha}_{d I^{PT}}^{v,\ell}+{\alpha}_{d K^{P}}^{v,\ell}\Big] (\bs{\mV}^\ell\times\langle\bs{J}\rangle)-\Big[{\alpha'}_{d I^{PT}}^{t,\ell}+{\alpha'}_{d K^{P}}^{t,\ell}\Big]{\mathcal{Q}}\cdot \dot{\bs{\mV}}^\ell\bigg]\\
\end{align*}

\section{Low Energy Limits of the Cosmic Field interactions} \label{sec: Low energy limits of the cosmic field interactions}
In the following, we derive the low-energy limits of the coupling between atoms and cosmic fields up to order $m^{-2}$. We treat the interaction as semiclassical. This allows us to derive the nonrelativistic interaction with the cosmic fields analogous to the derivation for the electromagnetic field in Refs.~\cite{bethe2012quantum, sakurai1967advanced}. 
  
For better transparency, we treat the scalar, pseudoscalar, vector, axial vector, and tensor components of $\Xi_{\mu\nu}$ all separately. We give a detailed derivation of the axial vector coupling. This introduces all the important ideas that are needed to derive the other four types.

% \subsection{Gamma Matrices}\label{sec: Gamma Matrices}
% We use the standard Dirac representation of the Gamma matrices:
% \begin{align*}
% & \gamma_{i}=\left(\begin{array}{cc}
% 0 & \sigma_i \\
% -\sigma_i & 0
% \end{array}\right),\quad \gamma_0=\left(\begin{array}{cc}
% 1 & 0 \\
% 0 & -1
% \end{array}\right)=\beta,\quad \alpha_i=\left(\begin{array}{cc}
% 0 & \sigma_i \\
% \sigma_i & 0
% \end{array}\right) \\
% & \gamma^5=\left(\begin{array}{cc}
% 0 & 1 \\
% 1 & 0
% \end{array}\right),\quad \Sigma_i=\left(\begin{array}{cc}
% \sigma_i & 0 \\
% 0 & \sigma_i
% \end{array}\right)
% \end{align*}
% where $\sigma_i$ are the Pauli matrices and $i\in\{1,2,3\}$.

\subsection{Axial Vector Coupling} \label{sec: axial vector coupling}
The Lagrangian for a fermion with an axial coupling to a cosmic field is given by:
\begin{align*}
  \mathcal{L}=i\bar{\psi}\gamma^\mu\partial_\mu\psi-m\bar{\psi}\psi-g_{Z'}\bar{\psi}\gamma^\mu\gamma^5Z'_\mu\psi
\end{align*}
Through the Euler-Lagrange equation, this Lagrangian implies the following equation of motion:
\begin{align*}
  -i\gamma^\mu\partial_\mu\psi+m\psi+g_{Z'}\gamma^\mu\gamma^5Z'_\mu\psi=0
\end{align*}
Through a multiplication by $\gamma^0$ (from the left side), we can separate spatial and temporal components to arrive at a relativistic Hamiltonian:
\begin{align}
  H\psi=(H^0+V_{Z'})\psi=i\partial_t\psi=\big(\bs{\alpha}\cdot\bs{p}+\beta m+g_{Z'}\gamma^5 Z'_0+g_{Z'}\bs{\Sigma}\cdot\bs{Z}'\big)\psi\label{eq: H psi Z}
\end{align}
Here $H^0$defines the free fermion's kinetic and potential energy, while $V_{Z'}$ characterizes the axial vector coupling to the cosmic field.

At this point, it is also possible to introduce additional potentials to model the system. To demonstrate this idea, we choose the simple example of the potential of a one-electron atom in the central field approximation $V_\Phi=-e\Phi(\bs{r})$. 
If one wants to investigate the interaction between cosmic fields and the electromagnetic field inside a medium, one could, also at this point, perform the minimal substitution $\bs{p}\rightarrow\bs{p}-e\bs{\mathcal{A}}$. However, to keep the treatment simple, we assume that such electromagnetic fields are absent in our system.

We now want to derive the low-energy form of the interaction potential $V_{Z'}$. For this, we can express Eq.~(\ref{eq: H psi Z}) as two coupled equations of the upper and lower components of the four-spinor $\psi=(\psi_{_U},\psi_{_L})^T$:
\begin{align*}
  \big[i\partial_t-m-g_{Z'}\bs{\sigma}\cdot\bs{Z}'+e\Phi\big]\psi_{_U}&=\big[g_{Z'}Z'_0+\bs{\sigma}\cdot\bs{p}\big]\psi_{_L}\\
  \big[i\partial_t+m-g_{Z'}\bs{\sigma}\cdot\bs{Z}'+e\Phi\big]\psi_{_L}&=\big[g_{Z'}Z'_0+\bs{\sigma}\cdot\bs{p}\big]\psi_{_U}
\end{align*}
In the non relativistic limit, the mass term dominates over all other. It is helpful to factor the wavefunction into a mass and a non relativistic component: $\psi=\psi^\text{NR}e^{-imt}$. Inserting this expression into the coupled equations above, leads to new coupled equations for the non relativistic components:
\begin{align*}
  \big[i\partial_t-g_{Z'}\bs{\sigma}\cdot\bs{Z}'+e\Phi\big]\psi_{_U}^\text{NR}&=\big[\bs{\sigma}\cdot\bs{p}+g_{Z'}Z'_0\big]\psi_{_L}^\text{NR}\\
  \big[i\partial_t+2m-g_{Z'}\bs{\sigma}\cdot\bs{Z}'+e\Phi\big]\psi_{_L}^\text{NR}&=\big[\bs{\sigma}\cdot\bs{p}+g_{Z'}Z'_0\big]\psi_{_U}^\text{NR}
\end{align*}
To obtain an equation for just the upper component, we can plug the second equation into the first:
\begin{align*}
  i\partial_t\psi_{_U}^\text{NR}=&\big[g_{Z'}\bs{\sigma}\cdot\bs{Z}'-e\Phi\big]\psi_{_U}^\text{NR}\\
  +&\big[\bs{\sigma}\cdot\bs{p}+g_{Z'}Z'_0\big]\big[i\partial_t+e\Phi-g_{Z'}\bs{\sigma}\cdot\bs{Z}'+2m\big]^{-1}\big[\bs{\sigma}\cdot\bs{p}+g_{Z'}Z'_0\big]\psi_{_U}^\text{NR}
\end{align*}
Because $m$ dominates all other contributions to the energy, we can perform the following Taylor expansion:
 \begin{align*}
   \frac{1}{i\partial_t +e\Psi-g_{Z'}\bs{\sigma}\cdot\bs{Z}'+2m}\simeq\frac{1}{2m}\bigg[1-\frac{i\partial_t+e\Phi-g_{Z'}\bs{\sigma}\cdot\bs{Z}'}{2m}\bigg]+\mathcal{O}(m^{-3})
 \end{align*}
In this approximation, we get:
\begin{align*}
  i\partial_t\psi_{_U}^\text{NR}=\bigg[&g_{Z'}\bs{\sigma}\cdot\bs{Z}'-e\Phi+\frac{p^2}{2m}+\frac{g_{Z'}}{2m}\{Z'_0,\bs{\sigma}\cdot\bs{p}\}-i\frac{p^2}{4m^2}\partial_t\\
  &-\frac{ig_{Z'}}{4m^2}\{ Z'_0,(\bs{\sigma}\cdot\bs{p})\partial_t\}-\frac{e}{4m^2}(\bs{\sigma}\cdot\bs{p})\Phi(\bs{\sigma}\cdot\bs{p})\\
  &-\frac{eg_{Z'}}{4m^2}\{Z'_0\Phi,\bs{\sigma}\cdot\bs{p}\}+\frac{g_{Z'}}{4m^2}(\bs{\sigma}\cdot\bs{p})(\bs{\sigma}\cdot\bs{Z}')(\bs{\sigma}\cdot\bs{p})\bigg]\psi_{_U}^\text{NR}
\end{align*}
where we already removed terms quadratic in $g_{Z'}$.

The terms independent of $Z'_\mu$ describe the Schrödinger Hamiltonian with relativistic corrections. In the following, we are only interested in the new terms that contain couplings to $Z'_\mu$. When a relativistic correction to an already existing term appears (e.g. $[1+\tfrac{e\Phi}{2m}]$ or $[1+\tfrac{\partial_t}{2m}]$), we ignore the small correction factor. We can further use the identity $\bs{p}=-i\bs{\nabla}$, and express the anticommutator as $\{\bs{\nabla},\bs{Z}'\}=(\bs{\nabla}\cdot\bs{Z}')+2\bs{Z}'\cdot\bs{\nabla}$, where the parenthesis indicates that the derivative only acts on $\bs{Z}'$. Under all of these considerations, we obtain the following potential to describe the interaction between the atom and the cosmic field:
\begin{align*}
  \frac{V_{Z'}}{g_{Z'}}=&\bs{\sigma}\cdot\bs{Z}'+\frac{1}{m}Z'_0(\bs{\sigma}\cdot\bs{p})-\frac{i}{2m}(\bs{\nabla}Z'_0)\cdot\bs{\sigma}-\frac{i}{4m^2}\dot{Z}'_0(\bs{\sigma\cdot\bs{p}})\\
  -&\frac{1}{4m^2}(\bs{\nabla}\dot{Z}'_0)\cdot\bs{\sigma}-\frac{e}{4m^2}\bs{\sigma}\cdot(\bs{\nabla}\Phi)Z'_0+\frac{1}{4m^2}(\bs{\sigma}\cdot\bs{p})(\bs{\sigma}\cdot\bs{Z}')(\bs{\sigma}\cdot\bs{p})
\end{align*}
We can see that the interaction between the Coulomb potential $V_\Phi=-e\Phi(\bs{r})$ and the cosmic field only appears in order $m^{-2}$. It can be shown that this is generally true for all operators that exist in the nonrelativistic limit and for all five types of cosmic fields. The quantity $\bs{\nabla}\Phi$ in particular necessarily vanishes inside an atomic system due to Schiff's Theorem \cite{schiff1963measurability}. For these reasons, we will not discuss direct couplings to interatomic potentials in the remainder of this paper.

The expression $(\bs{\sigma}\cdot\bs{p})(\bs{\sigma}\cdot\bs{Z}')(\bs{\sigma}\cdot\bs{p})$ can further be simplified. For this, we can use the defining algebraic relation of the Pauli matrices: $\sigma_i\sigma_j=\delta_{ij}+i\varepsilon_{ijk}\sigma_k$. From this, we get:
\begin{align*}
  (\bs{\sigma}\cdot\bs{p})(\bs{\sigma}\cdot\bs{Z}')(\bs{\sigma}\cdot\bs{p})=&2\bs{Z}'\cdot\bs{p}(\bs{\sigma\cdot\bs{p}})-i(\bs{\nabla}\cdot\bs{Z}')(\bs{\sigma}\cdot\bs{p}) +(\bs{\nabla}\times\bs{Z}')\cdot\bs{\sigma}(\bs{\sigma}\cdot\bs{p})
\end{align*}
We arrive at a final expression for the nonrelativistic approximation of the axial vector interaction of the atom and the cosmic field:
\begin{align*}
  \frac{V_{Z'}}{g_{Z'}}=&\bs{\sigma}\cdot\bs{Z}'-\frac{1}{m}Z'_0(\bs{\sigma}\cdot\bs{p})-\frac{i}{2m}(\bs{\nabla}Z'_0)\cdot\bs{\sigma}-\frac{i}{4m^2}\dot{Z}'_0(\bs{\sigma\cdot\bs{p}})\\
  -&\frac{1}{4m^2}(\bs{\nabla}\dot{Z}'_0)\cdot\bs{\sigma}-\frac{i}{4m^2}(\bs{\nabla}\cdot\bs{Z}')(\bs{\sigma}\cdot\bs{p})\\
  +&\frac{1}{4m^2}(\bs{\nabla}\times\bs{Z}')\cdot\bs{\sigma}(\bs{\sigma}\cdot\bs{p})+\frac{1}{2m^2}\bs{Z}'\cdot\bs{p}(\bs{\sigma\cdot\bs{p}})
\end{align*}
\subsection{Pseudoscalar Coupling}
The derivation for a pseudoscalar (axion) coupling can be performed analogously to the axial vector coupling. The Lagrangian $\mathcal{L}_a=-ig_a\bar{\psi}a\gamma^5\psi$ implies the following equation for the (relativistic) Dirac spinor $\psi$:
\begin{align*}
  i\partial_t\psi=(ig_a\beta\gamma^5a+\bs{\alpha}\cdot\bs{p}+\beta m)\psi
\end{align*}
This, in turn, implies the following equation for the nonrelativistic upper component (Pauli Spinor) $\psi^\text{NR}_{_U} $:
\begin{align*}
   i\partial_t\psi_{_U}^\text{NR}=\big[\bs{\sigma}\cdot\bs{p}+ig_a a\big]\big[i\partial_t+2m\big]^{-1}\big[\bs{\sigma}\cdot\bs{p}-ig_a a\big]\psi_{_U}^\text{NR}
\end{align*}
In the nonrelativistic limit up to order $m^{-2}$, we obtain the following potential for the pseudoscalar coupling of an atom to a cosmic field:
\begin{align*}
  \frac{V_a}{g_a}=-\frac{1}{2m}\bs{\sigma}\cdot(\bs{\nabla}{a})+\frac{i}{4m^2}\bs{\sigma}\cdot(\bs{\nabla}\dot{a})+\frac{1}{4m^2}\dot{a}(\bs{\sigma\cdot\bs{p}})
\end{align*}
The coupling to the axion is often expressed as $\mathcal{L}_a=\bar{\psi}\gamma^5\gamma^\mu\partial_\mu a\psi$. In Refs.~\cite{smith2024fermionic, quevillon2019axions} it was argued that this derivative axial vector coupling is equivalent to the pseudoscalar coupling discussed here.

\subsection{Vector Coupling}
The Lagrangian of the vector coupling ($\mathcal{L}_{A'}=-g_{A'}\bar{\psi}\gamma^\mu A'_\mu\psi$) implies the following equation for $\psi$:
\begin{align*}
  i\partial_t\psi=\big(\bs{\alpha}\cdot(\bs{p}-g_{A'}\bs{A}')+g_{A'}A_0'+\beta m\big)\psi
\end{align*}
For the Pauli Spinor $\psi^\text{NR}_{_U} $, this implies:
\begin{align*}
   i\partial_t\psi_{_U}^\text{NR}=g_{A'}A'_0\psi_{_U}^\text{NR}+\big[\bs{\sigma}\cdot\bs{p}-g_{A'}\bs{\sigma}\cdot \bs{A}'\big]\big[i\partial_t-g_{A'}A'_0+2m\big]^{-1}\big[\bs{\sigma}\cdot\bs{p}-g_{A'}\bs{\sigma}\cdot\bs{A}'\big]\psi_{_U}^\text{NR}
\end{align*}
In the non relativistic limit, this leads to the following potential:
\begin{align*}
  \frac{V_{A'}}{g_{A'}}=A'_0-\frac{1}{2m}\big\{\bs{\sigma}\cdot\bs{p},\bs{\sigma}\cdot\bs{A}'\big\}+\frac{i}{4m^2}\big\{(\bs{\sigma}\cdot\bs{p})\partial_t,\bs{\sigma}\cdot\bs{A}'\big\}+\frac{1}{4m^2}(\bs{\sigma}\cdot\bs{p})A'_0(\bs{\sigma}\cdot\bs{p})
\end{align*}
By using the algebraic relations of the Pauli matrices, we can solve the products of Pauli matrices:
\begin{align*}
    \frac{ V_{A'}}{g_{A'}}=&\Big(1+\frac{p^2}{4m^2}\Big)A'_0-\frac{1}{m}\bs{A}'\cdot\bs{p}+\frac{i}{2m}(\bs{\nabla}\cdot\bs{A}')-\frac{1}{2m}(\bs{\nabla}\times\bs{A}')\cdot\bs{\sigma}
\\
-&\frac{i}{4m^2}(\bs{\nabla}A'_0)\cdot\bs{\sigma}(\bs{\sigma\cdot\bs{p}})+\frac{1}{4m^2}(\bs{\nabla}\cdot\dot{\bs{A}}')\\
+&\frac{i}{4m^2}(\bs{\nabla}\times\dot{\bs{A}}')\cdot\bs{\sigma}
+\frac{i}{4m^2}\dot{\bs{A}}'\cdot\bs{p}-\frac{1}{4m^2}\dot{\bs{A}}'\cdot(\bs{\sigma}\times\bs{p})
\end{align*}
Here, the first term is the sum of a constant and a higher-order relativistic correction ($p^2/4m^2$). It is included because the constant potential can not lead to any physical observable. It can simply be removed by a redefinition of the potential energy. For the same reason, the terms $\frac{i}{2m}(\bs{\nabla}\cdot\bs{A}')$, and $\frac{1}{4m^2}(\bs{\nabla}\cdot\dot{\bs{A}}')$ are not measurable.
\subsection{Scalar Coupling}
The Lagrangian describing the scalar coupling has the very simple form: $\mathcal{L}_{\phi}=-g_\phi\bar{\psi}\phi\psi$. It leads to the following equation of motion:
\begin{align}
i\partial_t\psi=\big(\bs{\alpha}\cdot\bs{p}+g_{\phi}\beta\phi+\beta m\big)\psi \label{eq: eom scalar}
\end{align}
For the upper component of the Pauli spinor follows:
\begin{align*}
  i\partial_t\psi_{_U}^\text{NR}=g_\phi\phi\psi_{_U}^\text{NR}+\big[\bs{\sigma}\cdot\bs{p}\big]\big[i\partial_t+g_\phi\phi+2m\big]^{-1}\big[\bs{\sigma}\cdot\bs{p}\big]\psi_{_U}^\text{NR}
\end{align*}
Up to order $m^{-2}$, this gives the following potential for the scalar coupling of an atom to a cosmic field:
\begin{align}
  \frac{V_\phi}{g_\phi}=\Big(1-\frac{p^2}{4m^2}\Big)\phi-\frac{i}{4m^2}(\bs{\nabla}\phi)\cdot\bs{\sigma}(\bs{\sigma}\cdot\bs{p})
\label{eq:Vphi_app}
\end{align}
Like before for $\bs{A}'$, the first term can be removed by a redefinition of the potential.
\subsection{Tensor Coupling} \label{sec: tensor coupling}
The Lagrangian of the tensor coupling is $\mathcal{L}_{\Theta}=-g_\Theta\bar{\psi}\sigma_{\mu\nu}\Theta^{\mu\nu}\psi$. This implies the following equation for $\psi$:
\begin{align*}
  i\partial_t\psi=\big(\bs{\alpha}\cdot\bs{p}+ig_\Theta \beta\alpha_i\big[\Theta_{0i}-\Theta_{i0}\big]+g_\Theta\varepsilon_{ij l }\Theta_{ij}\beta\Sigma_ l +\beta m\big)\psi
\end{align*}
It is apparent that only antisymmetric components of $\Theta_{\mu\nu}$ can have an influence on the system. We can define the two vectors $\theta^E_i=\Theta_{0i}$, and $\theta_ l ^B=-\tfrac{1}{2}\varepsilon_{ij l }\Theta_{ij}$ in analogy to the electric and magnetic field components of the electromagnetic field tensor \cite{bethe2012quantum}. This yields an equation that is mathematically equivalent to the anomalous magnetic dipole moment \cite{roberts2010lepton}:
\begin{align*}
  i\partial_t\psi=\big(\bs{\alpha}\cdot\bs{p}+2ig_\Theta \beta\bs{\alpha}\cdot\bs{\theta}^E-2g_\Theta\beta\bs{\Sigma}\cdot\bs{\theta}^B+\beta m\big)\psi
\end{align*}
For the non relativistic wavefunction $\psi^{NR}_U$, this implies:
\begin{align*}
  i\partial_t\psi_{_U}^\text{NR}=&-2g_\phi\big[\bs{\sigma\cdot\bs{\theta}^B}\big]\psi_{_U}^\text{NR}\\
  +&\big[\bs{\sigma}\cdot\bs{p}+2ig_\Theta\bs{\sigma}\cdot\bs{\theta}^E\big]\big[i\partial_t+2g_\phi\bs{\sigma\cdot\bs{\theta}^B}+2m\big]^{-1}\big[\bs{\sigma}\cdot\bs{p}-2ig_\Theta\bs{\sigma}\cdot\bs{\theta}^E\big]\psi_{_U}^\text{NR}
\end{align*}
Up to order $m^{-2}$, this gives the following potential for the tensor coupling of an atom to a cosmic field:
\begin{align*}
  \frac{V_\Theta}{g_\Theta}=&-2\bs{\sigma}\cdot\bs{\theta}^B-\frac{1}{m}(\bs{\nabla}\cdot\bs{\theta}^E)-\frac{i}{m}(\bs{\nabla}\times\bs{\theta}^E)\cdot\bs{\sigma}+\frac{2}{m}\bs{\theta}^E\cdot(\bs{\sigma}\times\bs{p})\\
  +&\frac{i}{2m^2}(\bs{\nabla}\cdot\dot{\bs{\theta}}^E)-\frac{1}{2m^2}(\bs{\nabla}\times\dot{\bs{\theta}}^E)\cdot\bs{\sigma}-\frac{1}{2m^2}\dot{\bs{\theta}}^E\cdot\bs{p}-\frac{i}{2m^2}\dot{\bs{\theta}}^E\cdot(\bs{\sigma}\times\bs{p})\\
  +&\frac{i}{2m^2}(\bs{\nabla}\cdot\bs{\theta}^B)(\bs{\sigma}\cdot\bs{p})-\frac{1}{2m^2}(\bs{\nabla}\times\bs{\theta}^B)\cdot\bs{\sigma}(\bs{\sigma}\cdot\bs{p})-\frac{1}{m^2}\bs{\theta}^B\cdot\bs{p}(\bs{\sigma}\cdot\bs{p})
\end{align*}
The terms $\frac{i}{2m^2}(\bs{\nabla}\cdot\dot{\bs{\theta}}^E)$ and $\frac{1}{m}(\bs{\nabla}\cdot\bs{\theta}^E)$ can again be removed as their atomic operators are simply constants.
\end{widetext}

\bibliography{References}

\begin{thebibliography}{75}%
\makeatletter
\providecommand \@ifxundefined [1]{%
 \@ifx{#1\undefined}
}%
\providecommand \@ifnum [1]{%
 \ifnum #1\expandafter \@firstoftwo
 \else \expandafter \@secondoftwo
 \fi
}%
\providecommand \@ifx [1]{%
 \ifx #1\expandafter \@firstoftwo
 \else \expandafter \@secondoftwo
 \fi
}%
\providecommand \natexlab [1]{#1}%
\providecommand \enquote  [1]{``#1''}%
\providecommand \bibnamefont  [1]{#1}%
\providecommand \bibfnamefont [1]{#1}%
\providecommand \citenamefont [1]{#1}%
\providecommand \href@noop [0]{\@secondoftwo}%
\providecommand \href [0]{\begingroup \@sanitize@url \@href}%
\providecommand \@href[1]{\@@startlink{#1}\@@href}%
\providecommand \@@href[1]{\endgroup#1\@@endlink}%
\providecommand \@sanitize@url [0]{\catcode `\\12\catcode `\$12\catcode `\&12\catcode `\#12\catcode `\^12\catcode `\_12\catcode `\%12\relax}%
\providecommand \@@startlink[1]{}%
\providecommand \@@endlink[0]{}%
\providecommand \url  [0]{\begingroup\@sanitize@url \@url }%
\providecommand \@url [1]{\endgroup\@href {#1}{\urlprefix }}%
\providecommand \urlprefix  [0]{URL }%
\providecommand \Eprint [0]{\href }%
\providecommand \doibase [0]{https://doi.org/}%
\providecommand \selectlanguage [0]{\@gobble}%
\providecommand \bibinfo  [0]{\@secondoftwo}%
\providecommand \bibfield  [0]{\@secondoftwo}%
\providecommand \translation [1]{[#1]}%
\providecommand \BibitemOpen [0]{}%
\providecommand \bibitemStop [0]{}%
\providecommand \bibitemNoStop [0]{.\EOS\space}%
\providecommand \EOS [0]{\spacefactor3000\relax}%
\providecommand \BibitemShut  [1]{\csname bibitem#1\endcsname}%
\let\auto@bib@innerbib\@empty
%</preamble>
\bibitem [{\citenamefont {Preskill}\ \emph {et~al.}(1983)\citenamefont {Preskill}, \citenamefont {Wise},\ and\ \citenamefont {Wilczek}}]{preskill1983cosmology}%
  \BibitemOpen
  \bibfield  {author} {\bibinfo {author} {\bibfnamefont {J.}~\bibnamefont {Preskill}}, \bibinfo {author} {\bibfnamefont {M.~B.}\ \bibnamefont {Wise}},\ and\ \bibinfo {author} {\bibfnamefont {F.}~\bibnamefont {Wilczek}},\ }\bibfield  {title} {\bibinfo {title} {Cosmology of the invisible axion},\ }\href {https://doi.org/10.1016/0370-2693(83)90637-8} {\bibfield  {journal} {\bibinfo  {journal} {Physics Letters B}\ }\textbf {\bibinfo {volume} {120}},\ \bibinfo {pages} {127} (\bibinfo {year} {1983})}\BibitemShut {NoStop}%
\bibitem [{\citenamefont {Ratra}\ and\ \citenamefont {Peebles}(1988)}]{ratra1988cosmological}%
  \BibitemOpen
  \bibfield  {author} {\bibinfo {author} {\bibfnamefont {B.}~\bibnamefont {Ratra}}\ and\ \bibinfo {author} {\bibfnamefont {P.~J.}\ \bibnamefont {Peebles}},\ }\bibfield  {title} {\bibinfo {title} {Cosmological consequences of a rolling homogeneous scalar field},\ }\href@noop {} {\bibfield  {journal} {\bibinfo  {journal} {Physical Review D}\ }\textbf {\bibinfo {volume} {37}},\ \bibinfo {pages} {3406} (\bibinfo {year} {1988})}\BibitemShut {NoStop}%
\bibitem [{\citenamefont {Graham}\ \emph {et~al.}(2015{\natexlab{a}})\citenamefont {Graham}, \citenamefont {Kaplan},\ and\ \citenamefont {Rajendran}}]{graham2015cosmological}%
  \BibitemOpen
  \bibfield  {author} {\bibinfo {author} {\bibfnamefont {P.~W.}\ \bibnamefont {Graham}}, \bibinfo {author} {\bibfnamefont {D.~E.}\ \bibnamefont {Kaplan}},\ and\ \bibinfo {author} {\bibfnamefont {S.}~\bibnamefont {Rajendran}},\ }\bibfield  {title} {\bibinfo {title} {Cosmological relaxation of the electroweak scale},\ }\href@noop {} {\bibfield  {journal} {\bibinfo  {journal} {Physical review letters}\ }\textbf {\bibinfo {volume} {115}},\ \bibinfo {pages} {221801} (\bibinfo {year} {2015}{\natexlab{a}})}\BibitemShut {NoStop}%
\bibitem [{\citenamefont {Weinberg}(1978)}]{weinberg1978new}%
  \BibitemOpen
  \bibfield  {author} {\bibinfo {author} {\bibfnamefont {S.}~\bibnamefont {Weinberg}},\ }\bibfield  {title} {\bibinfo {title} {A new light boson?},\ }\href@noop {} {\bibfield  {journal} {\bibinfo  {journal} {Physical Review Letters}\ }\textbf {\bibinfo {volume} {40}},\ \bibinfo {pages} {223} (\bibinfo {year} {1978})}\BibitemShut {NoStop}%
\bibitem [{\citenamefont {Wilczek}(1978{\natexlab{a}})}]{wilczek1978problem}%
  \BibitemOpen
  \bibfield  {author} {\bibinfo {author} {\bibfnamefont {F.}~\bibnamefont {Wilczek}},\ }\bibfield  {title} {\bibinfo {title} {Problem of strong {P} and {T} invariance in the presence of instantons},\ }\href@noop {} {\bibfield  {journal} {\bibinfo  {journal} {Physical Review Letters}\ }\textbf {\bibinfo {volume} {40}},\ \bibinfo {pages} {279} (\bibinfo {year} {1978}{\natexlab{a}})}\BibitemShut {NoStop}%
\bibitem [{\citenamefont {Kim}(1979)}]{kim1979weak}%
  \BibitemOpen
  \bibfield  {author} {\bibinfo {author} {\bibfnamefont {J.~E.}\ \bibnamefont {Kim}},\ }\bibfield  {title} {\bibinfo {title} {Weak-interaction singlet and strong {CP} invariance},\ }\href@noop {} {\bibfield  {journal} {\bibinfo  {journal} {Physical Review Letters}\ }\textbf {\bibinfo {volume} {43}},\ \bibinfo {pages} {103} (\bibinfo {year} {1979})}\BibitemShut {NoStop}%
\bibitem [{\citenamefont {Fritzsch}\ and\ \citenamefont {Minkowski}(1975)}]{fritzsch1975unified}%
  \BibitemOpen
  \bibfield  {author} {\bibinfo {author} {\bibfnamefont {H.}~\bibnamefont {Fritzsch}}\ and\ \bibinfo {author} {\bibfnamefont {P.}~\bibnamefont {Minkowski}},\ }\bibfield  {title} {\bibinfo {title} {Unified interactions of leptons and hadrons},\ }\href@noop {} {\bibfield  {journal} {\bibinfo  {journal} {Annals of Physics}\ }\textbf {\bibinfo {volume} {93}},\ \bibinfo {pages} {193} (\bibinfo {year} {1975})}\BibitemShut {NoStop}%
\bibitem [{\citenamefont {Witten}(1980)}]{witten1980neutrino}%
  \BibitemOpen
  \bibfield  {author} {\bibinfo {author} {\bibfnamefont {E.}~\bibnamefont {Witten}},\ }\bibfield  {title} {\bibinfo {title} {Neutrino masses in the minimal o (10) theory},\ }\href@noop {} {\bibfield  {journal} {\bibinfo  {journal} {Physics Letters B}\ }\textbf {\bibinfo {volume} {91}},\ \bibinfo {pages} {81} (\bibinfo {year} {1980})}\BibitemShut {NoStop}%
\bibitem [{\citenamefont {Witten}(1984)}]{witten1984some}%
  \BibitemOpen
  \bibfield  {author} {\bibinfo {author} {\bibfnamefont {E.}~\bibnamefont {Witten}},\ }\bibfield  {title} {\bibinfo {title} {Some properties of o (32) superstrings},\ }\href@noop {} {\bibfield  {journal} {\bibinfo  {journal} {Physics Letters B}\ }\textbf {\bibinfo {volume} {149}},\ \bibinfo {pages} {351} (\bibinfo {year} {1984})}\BibitemShut {NoStop}%
\bibitem [{\citenamefont {Svrcek}\ and\ \citenamefont {Witten}(2006)}]{svrcek2006axions}%
  \BibitemOpen
  \bibfield  {author} {\bibinfo {author} {\bibfnamefont {P.}~\bibnamefont {Svrcek}}\ and\ \bibinfo {author} {\bibfnamefont {E.}~\bibnamefont {Witten}},\ }\bibfield  {title} {\bibinfo {title} {Axions in string theory},\ }\href@noop {} {\bibfield  {journal} {\bibinfo  {journal} {Journal of High Energy Physics}\ }\textbf {\bibinfo {volume} {2006}},\ \bibinfo {pages} {051} (\bibinfo {year} {2006})}\BibitemShut {NoStop}%
\bibitem [{\citenamefont {Goodsell}\ \emph {et~al.}(2009)\citenamefont {Goodsell}, \citenamefont {Jaeckel}, \citenamefont {Redondo},\ and\ \citenamefont {Ringwald}}]{goodsell2009naturally}%
  \BibitemOpen
  \bibfield  {author} {\bibinfo {author} {\bibfnamefont {M.}~\bibnamefont {Goodsell}}, \bibinfo {author} {\bibfnamefont {J.}~\bibnamefont {Jaeckel}}, \bibinfo {author} {\bibfnamefont {J.}~\bibnamefont {Redondo}},\ and\ \bibinfo {author} {\bibfnamefont {A.}~\bibnamefont {Ringwald}},\ }\bibfield  {title} {\bibinfo {title} {Naturally light hidden photons in large volume string compactifications},\ }\href@noop {} {\bibfield  {journal} {\bibinfo  {journal} {Journal of High Energy Physics}\ }\textbf {\bibinfo {volume} {2009}},\ \bibinfo {pages} {027} (\bibinfo {year} {2009})}\BibitemShut {NoStop}%
\bibitem [{\citenamefont {Witten}(1981)}]{witten1981search}%
  \BibitemOpen
  \bibfield  {author} {\bibinfo {author} {\bibfnamefont {E.}~\bibnamefont {Witten}},\ }\bibfield  {title} {\bibinfo {title} {Search for a realistic kaluza-klein theory},\ }\href@noop {} {\bibfield  {journal} {\bibinfo  {journal} {Nuclear Physics B}\ }\textbf {\bibinfo {volume} {186}},\ \bibinfo {pages} {412} (\bibinfo {year} {1981})}\BibitemShut {NoStop}%
\bibitem [{\citenamefont {Kostelecky}\ and\ \citenamefont {Lane}(1999)}]{kostelecky1999nonrelativistic}%
  \BibitemOpen
  \bibfield  {author} {\bibinfo {author} {\bibfnamefont {A.}~\bibnamefont {Kostelecky}}\ and\ \bibinfo {author} {\bibfnamefont {C.}~\bibnamefont {Lane}},\ }\bibfield  {title} {\bibinfo {title} {Nonrelativistic quantum hamiltonian for lorentz violation},\ }\href@noop {} {\bibfield  {journal} {\bibinfo  {journal} {arXiv preprint hep-ph/9909542}\ } (\bibinfo {year} {1999})}\BibitemShut {NoStop}%
\bibitem [{\citenamefont {Safronova}(2023)}]{safronova2023searches}%
  \BibitemOpen
  \bibfield  {author} {\bibinfo {author} {\bibfnamefont {M.~S.}\ \bibnamefont {Safronova}},\ }\bibfield  {title} {\bibinfo {title} {Searches for new physics},\ }in\ \href@noop {} {\emph {\bibinfo {booktitle} {Springer Handbook of Atomic, Molecular, and Optical Physics}}}\ (\bibinfo  {publisher} {Springer},\ \bibinfo {year} {2023})\ pp.\ \bibinfo {pages} {471--484}\BibitemShut {NoStop}%
\bibitem [{\citenamefont {O’Hare}\ \emph {et~al.}(2025)\citenamefont {O’Hare} \emph {et~al.}}]{o2025cajohare}%
  \BibitemOpen
  \bibfield  {author} {\bibinfo {author} {\bibfnamefont {C.}~\bibnamefont {O’Hare}} \emph {et~al.},\ }\href {https://github.com/cajohare/AxionLimits} {\bibinfo {title} {https://github.com/cajohare/axionlimits}} (\bibinfo {year} {2025})\BibitemShut {NoStop}%
\bibitem [{\citenamefont {Roberts}\ \emph {et~al.}(2014{\natexlab{a}})\citenamefont {Roberts}, \citenamefont {Stadnik}, \citenamefont {Dzuba}, \citenamefont {Flambaum}, \citenamefont {Leefer},\ and\ \citenamefont {Budker}}]{roberts2014limiting}%
  \BibitemOpen
  \bibfield  {author} {\bibinfo {author} {\bibfnamefont {B.}~\bibnamefont {Roberts}}, \bibinfo {author} {\bibfnamefont {Y.}~\bibnamefont {Stadnik}}, \bibinfo {author} {\bibfnamefont {V.}~\bibnamefont {Dzuba}}, \bibinfo {author} {\bibfnamefont {V.}~\bibnamefont {Flambaum}}, \bibinfo {author} {\bibfnamefont {N.}~\bibnamefont {Leefer}},\ and\ \bibinfo {author} {\bibfnamefont {D.}~\bibnamefont {Budker}},\ }\bibfield  {title} {\bibinfo {title} {Limiting p-odd interactions of cosmic fields with electrons, protons, and neutrons},\ }\href@noop {} {\bibfield  {journal} {\bibinfo  {journal} {Physical Review Letters}\ }\textbf {\bibinfo {volume} {113}},\ \bibinfo {pages} {081601} (\bibinfo {year} {2014}{\natexlab{a}})}\BibitemShut {NoStop}%
\bibitem [{\citenamefont {Roberts}\ \emph {et~al.}(2014{\natexlab{b}})\citenamefont {Roberts}, \citenamefont {Stadnik}, \citenamefont {Dzuba}, \citenamefont {Flambaum}, \citenamefont {Leefer},\ and\ \citenamefont {Budker}}]{roberts2014parity}%
  \BibitemOpen
  \bibfield  {author} {\bibinfo {author} {\bibfnamefont {B.}~\bibnamefont {Roberts}}, \bibinfo {author} {\bibfnamefont {Y.}~\bibnamefont {Stadnik}}, \bibinfo {author} {\bibfnamefont {V.}~\bibnamefont {Dzuba}}, \bibinfo {author} {\bibfnamefont {V.}~\bibnamefont {Flambaum}}, \bibinfo {author} {\bibfnamefont {N.}~\bibnamefont {Leefer}},\ and\ \bibinfo {author} {\bibfnamefont {D.}~\bibnamefont {Budker}},\ }\bibfield  {title} {\bibinfo {title} {Parity-violating interactions of cosmic fields with atoms, molecules, and nuclei: Concepts and calculations for laboratory searches and extracting limits},\ }\href@noop {} {\bibfield  {journal} {\bibinfo  {journal} {Physical Review D}\ }\textbf {\bibinfo {volume} {90}},\ \bibinfo {pages} {096005} (\bibinfo {year} {2014}{\natexlab{b}})}\BibitemShut {NoStop}%
\bibitem [{\citenamefont {Gaul}\ \emph {et~al.}(2020)\citenamefont {Gaul}, \citenamefont {Kozlov}, \citenamefont {Isaev},\ and\ \citenamefont {Berger}}]{gaul2020chiral}%
  \BibitemOpen
  \bibfield  {author} {\bibinfo {author} {\bibfnamefont {K.}~\bibnamefont {Gaul}}, \bibinfo {author} {\bibfnamefont {M.~G.}\ \bibnamefont {Kozlov}}, \bibinfo {author} {\bibfnamefont {T.~A.}\ \bibnamefont {Isaev}},\ and\ \bibinfo {author} {\bibfnamefont {R.}~\bibnamefont {Berger}},\ }\bibfield  {title} {\bibinfo {title} {Chiral molecules as sensitive probes for direct detection of p-odd cosmic fields},\ }\href@noop {} {\bibfield  {journal} {\bibinfo  {journal} {Physical review letters}\ }\textbf {\bibinfo {volume} {125}},\ \bibinfo {pages} {123004} (\bibinfo {year} {2020})}\BibitemShut {NoStop}%
\bibitem [{\citenamefont {Berlin}\ \emph {et~al.}(2024)\citenamefont {Berlin}, \citenamefont {Millar}, \citenamefont {Trickle},\ and\ \citenamefont {Zhou}}]{berlin2024physical}%
  \BibitemOpen
  \bibfield  {author} {\bibinfo {author} {\bibfnamefont {A.}~\bibnamefont {Berlin}}, \bibinfo {author} {\bibfnamefont {A.~J.}\ \bibnamefont {Millar}}, \bibinfo {author} {\bibfnamefont {T.}~\bibnamefont {Trickle}},\ and\ \bibinfo {author} {\bibfnamefont {K.}~\bibnamefont {Zhou}},\ }\bibfield  {title} {\bibinfo {title} {Physical signatures of fermion-coupled axion dark matter},\ }\href@noop {} {\bibfield  {journal} {\bibinfo  {journal} {Journal of High Energy Physics}\ }\textbf {\bibinfo {volume} {2024}},\ \bibinfo {pages} {1} (\bibinfo {year} {2024})}\BibitemShut {NoStop}%
\bibitem [{\citenamefont {Smith}(2024)}]{smith2024fermionic}%
  \BibitemOpen
  \bibfield  {author} {\bibinfo {author} {\bibfnamefont {C.}~\bibnamefont {Smith}},\ }\bibfield  {title} {\bibinfo {title} {On the fermionic couplings of axionic dark matter},\ }\href@noop {} {\bibfield  {journal} {\bibinfo  {journal} {The European Physical Journal C}\ }\textbf {\bibinfo {volume} {84}},\ \bibinfo {pages} {12} (\bibinfo {year} {2024})}\BibitemShut {NoStop}%
\bibitem [{\citenamefont {Peskin}(2018)}]{peskin2018introduction}%
  \BibitemOpen
  \bibfield  {author} {\bibinfo {author} {\bibfnamefont {M.~E.}\ \bibnamefont {Peskin}},\ }\href@noop {} {\emph {\bibinfo {title} {An Introduction to quantum field theory}}}\ (\bibinfo  {publisher} {CRC press},\ \bibinfo {year} {2018})\BibitemShut {NoStop}%
\bibitem [{\citenamefont {Wilczek}(1978{\natexlab{b}})}]{Wilczek:1977pj}%
  \BibitemOpen
  \bibfield  {author} {\bibinfo {author} {\bibfnamefont {F.}~\bibnamefont {Wilczek}},\ }\bibfield  {title} {\bibinfo {title} {{Problem of Strong $P$ and $T$ Invariance in the Presence of Instantons}},\ }\href {https://doi.org/10.1103/PhysRevLett.40.279} {\bibfield  {journal} {\bibinfo  {journal} {Phys. Rev. Lett.}\ }\textbf {\bibinfo {volume} {40}},\ \bibinfo {pages} {279} (\bibinfo {year} {1978}{\natexlab{b}})}\BibitemShut {NoStop}%
\bibitem [{\citenamefont {Kim}\ and\ \citenamefont {Carosi}(2010)}]{kim2010axions}%
  \BibitemOpen
  \bibfield  {author} {\bibinfo {author} {\bibfnamefont {J.~E.}\ \bibnamefont {Kim}}\ and\ \bibinfo {author} {\bibfnamefont {G.}~\bibnamefont {Carosi}},\ }\bibfield  {title} {\bibinfo {title} {Axions and the strong {{CP}} problem},\ }\href {https://doi.org/10.1103/RevModPhys.82.557} {\bibfield  {journal} {\bibinfo  {journal} {Reviews of Modern Physics}\ }\textbf {\bibinfo {volume} {82}},\ \bibinfo {pages} {557} (\bibinfo {year} {2010})},\ \Eprint {https://arxiv.org/abs/0807.3125} {arXiv:0807.3125} \BibitemShut {NoStop}%
\bibitem [{\citenamefont {Graham}\ \emph {et~al.}(2015{\natexlab{b}})\citenamefont {Graham}, \citenamefont {Irastorza}, \citenamefont {Lamoreaux}, \citenamefont {Lindner},\ and\ \citenamefont {van Bibber}}]{graham2015experimental}%
  \BibitemOpen
  \bibfield  {author} {\bibinfo {author} {\bibfnamefont {P.~W.}\ \bibnamefont {Graham}}, \bibinfo {author} {\bibfnamefont {I.~G.}\ \bibnamefont {Irastorza}}, \bibinfo {author} {\bibfnamefont {S.~K.}\ \bibnamefont {Lamoreaux}}, \bibinfo {author} {\bibfnamefont {A.}~\bibnamefont {Lindner}},\ and\ \bibinfo {author} {\bibfnamefont {K.~A.}\ \bibnamefont {van Bibber}},\ }\bibfield  {title} {\bibinfo {title} {Experimental searches for the axion and axion-like particles},\ }\href@noop {} {\bibfield  {journal} {\bibinfo  {journal} {Annual Review of Nuclear and Particle Science}\ }\textbf {\bibinfo {volume} {65}},\ \bibinfo {pages} {485} (\bibinfo {year} {2015}{\natexlab{b}})}\BibitemShut {NoStop}%
\bibitem [{\citenamefont {Arkani-Hamed}\ \emph {et~al.}(2004)\citenamefont {Arkani-Hamed}, \citenamefont {Cheng}, \citenamefont {Luty},\ and\ \citenamefont {Mukohyama}}]{arkani2004ghost}%
  \BibitemOpen
  \bibfield  {author} {\bibinfo {author} {\bibfnamefont {N.}~\bibnamefont {Arkani-Hamed}}, \bibinfo {author} {\bibfnamefont {H.-C.}\ \bibnamefont {Cheng}}, \bibinfo {author} {\bibfnamefont {M.~A.}\ \bibnamefont {Luty}},\ and\ \bibinfo {author} {\bibfnamefont {S.}~\bibnamefont {Mukohyama}},\ }\bibfield  {title} {\bibinfo {title} {Ghost condensation and a consistent infrared modification of gravity},\ }\href@noop {} {\bibfield  {journal} {\bibinfo  {journal} {Journal of High Energy Physics}\ }\textbf {\bibinfo {volume} {2004}},\ \bibinfo {pages} {074} (\bibinfo {year} {2004})}\BibitemShut {NoStop}%
\bibitem [{\citenamefont {Capozziello}\ and\ \citenamefont {De~Laurentis}(2011)}]{capozziello2011extended}%
  \BibitemOpen
  \bibfield  {author} {\bibinfo {author} {\bibfnamefont {S.}~\bibnamefont {Capozziello}}\ and\ \bibinfo {author} {\bibfnamefont {M.}~\bibnamefont {De~Laurentis}},\ }\bibfield  {title} {\bibinfo {title} {Extended theories of gravity},\ }\href@noop {} {\bibfield  {journal} {\bibinfo  {journal} {Physics Reports}\ }\textbf {\bibinfo {volume} {509}},\ \bibinfo {pages} {167} (\bibinfo {year} {2011})}\BibitemShut {NoStop}%
\bibitem [{\citenamefont {Olive}\ and\ \citenamefont {Pospelov}(2008)}]{olive2008environmental}%
  \BibitemOpen
  \bibfield  {author} {\bibinfo {author} {\bibfnamefont {K.~A.}\ \bibnamefont {Olive}}\ and\ \bibinfo {author} {\bibfnamefont {M.}~\bibnamefont {Pospelov}},\ }\bibfield  {title} {\bibinfo {title} {Environmental dependence of masses and coupling constants},\ }\href@noop {} {\bibfield  {journal} {\bibinfo  {journal} {Physical Review D—Particles, Fields, Gravitation, and Cosmology}\ }\textbf {\bibinfo {volume} {77}},\ \bibinfo {pages} {043524} (\bibinfo {year} {2008})}\BibitemShut {NoStop}%
\bibitem [{\citenamefont {Linder}(2008)}]{linder2008dynamics}%
  \BibitemOpen
  \bibfield  {author} {\bibinfo {author} {\bibfnamefont {E.~V.}\ \bibnamefont {Linder}},\ }\bibfield  {title} {\bibinfo {title} {The dynamics of quintessence, the quintessence of dynamics},\ }\href@noop {} {\bibfield  {journal} {\bibinfo  {journal} {General Relativity and Gravitation}\ }\textbf {\bibinfo {volume} {40}},\ \bibinfo {pages} {329} (\bibinfo {year} {2008})}\BibitemShut {NoStop}%
\bibitem [{\citenamefont {Tsujikawa}(2013)}]{tsujikawa2013quintessence}%
  \BibitemOpen
  \bibfield  {author} {\bibinfo {author} {\bibfnamefont {S.}~\bibnamefont {Tsujikawa}},\ }\bibfield  {title} {\bibinfo {title} {Quintessence: a review},\ }\href@noop {} {\bibfield  {journal} {\bibinfo  {journal} {Classical and Quantum Gravity}\ }\textbf {\bibinfo {volume} {30}},\ \bibinfo {pages} {214003} (\bibinfo {year} {2013})}\BibitemShut {NoStop}%
\bibitem [{\citenamefont {Jackson~Kimball}\ and\ \citenamefont {Van~Bibber}(2023)}]{jackson2023search}%
  \BibitemOpen
  \bibfield  {author} {\bibinfo {author} {\bibfnamefont {D.~F.}\ \bibnamefont {Jackson~Kimball}}\ and\ \bibinfo {author} {\bibfnamefont {K.}~\bibnamefont {Van~Bibber}},\ }\href@noop {} {\emph {\bibinfo {title} {The search for ultralight bosonic dark matter}}}\ (\bibinfo  {publisher} {Springer Nature},\ \bibinfo {year} {2023})\BibitemShut {NoStop}%
\bibitem [{\citenamefont {Ferreira}(2021)}]{ferreira2021ultra}%
  \BibitemOpen
  \bibfield  {author} {\bibinfo {author} {\bibfnamefont {E.~G.}\ \bibnamefont {Ferreira}},\ }\bibfield  {title} {\bibinfo {title} {Ultra-light dark matter},\ }\href@noop {} {\bibfield  {journal} {\bibinfo  {journal} {The Astronomy and Astrophysics Review}\ }\textbf {\bibinfo {volume} {29}},\ \bibinfo {pages} {7} (\bibinfo {year} {2021})}\BibitemShut {NoStop}%
\bibitem [{\citenamefont {Arvanitaki}\ \emph {et~al.}(2015)\citenamefont {Arvanitaki}, \citenamefont {Huang},\ and\ \citenamefont {Van~Tilburg}}]{arvanitaki2015searching}%
  \BibitemOpen
  \bibfield  {author} {\bibinfo {author} {\bibfnamefont {A.}~\bibnamefont {Arvanitaki}}, \bibinfo {author} {\bibfnamefont {J.}~\bibnamefont {Huang}},\ and\ \bibinfo {author} {\bibfnamefont {K.}~\bibnamefont {Van~Tilburg}},\ }\bibfield  {title} {\bibinfo {title} {Searching for dilaton dark matter with atomic clocks},\ }\href@noop {} {\bibfield  {journal} {\bibinfo  {journal} {Physical Review D}\ }\textbf {\bibinfo {volume} {91}},\ \bibinfo {pages} {015015} (\bibinfo {year} {2015})}\BibitemShut {NoStop}%
\bibitem [{\citenamefont {Safronova}(2019)}]{safronova2019search}%
  \BibitemOpen
  \bibfield  {author} {\bibinfo {author} {\bibfnamefont {M.~S.}\ \bibnamefont {Safronova}},\ }\bibfield  {title} {\bibinfo {title} {The search for variation of fundamental constants with clocks},\ }\href@noop {} {\bibfield  {journal} {\bibinfo  {journal} {Annalen der Physik}\ }\textbf {\bibinfo {volume} {531}},\ \bibinfo {pages} {1800364} (\bibinfo {year} {2019})}\BibitemShut {NoStop}%
\bibitem [{\citenamefont {Cheng}\ \emph {et~al.}(2002)\citenamefont {Cheng}, \citenamefont {Feng},\ and\ \citenamefont {Matchev}}]{cheng2002kaluza}%
  \BibitemOpen
  \bibfield  {author} {\bibinfo {author} {\bibfnamefont {H.-C.}\ \bibnamefont {Cheng}}, \bibinfo {author} {\bibfnamefont {J.~L.}\ \bibnamefont {Feng}},\ and\ \bibinfo {author} {\bibfnamefont {K.~T.}\ \bibnamefont {Matchev}},\ }\bibfield  {title} {\bibinfo {title} {Kaluza-klein dark matter},\ }\href@noop {} {\bibfield  {journal} {\bibinfo  {journal} {Physical review letters}\ }\textbf {\bibinfo {volume} {89}},\ \bibinfo {pages} {211301} (\bibinfo {year} {2002})}\BibitemShut {NoStop}%
\bibitem [{\citenamefont {Abel}\ \emph {et~al.}(2008)\citenamefont {Abel}, \citenamefont {Goodsell}, \citenamefont {Jaeckel}, \citenamefont {Khoze},\ and\ \citenamefont {Ringwald}}]{abel2008kinetic}%
  \BibitemOpen
  \bibfield  {author} {\bibinfo {author} {\bibfnamefont {S.~A.}\ \bibnamefont {Abel}}, \bibinfo {author} {\bibfnamefont {M.~D.}\ \bibnamefont {Goodsell}}, \bibinfo {author} {\bibfnamefont {J.}~\bibnamefont {Jaeckel}}, \bibinfo {author} {\bibfnamefont {V.}~\bibnamefont {Khoze}},\ and\ \bibinfo {author} {\bibfnamefont {A.}~\bibnamefont {Ringwald}},\ }\bibfield  {title} {\bibinfo {title} {Kinetic mixing of the photon with hidden u (1) s in string phenomenology},\ }\href@noop {} {\bibfield  {journal} {\bibinfo  {journal} {Journal of High Energy Physics}\ }\textbf {\bibinfo {volume} {2008}},\ \bibinfo {pages} {124} (\bibinfo {year} {2008})}\BibitemShut {NoStop}%
\bibitem [{\citenamefont {Caputo}\ \emph {et~al.}(2021)\citenamefont {Caputo}, \citenamefont {Millar}, \citenamefont {O’Hare},\ and\ \citenamefont {Vitagliano}}]{caputo2021dark}%
  \BibitemOpen
  \bibfield  {author} {\bibinfo {author} {\bibfnamefont {A.}~\bibnamefont {Caputo}}, \bibinfo {author} {\bibfnamefont {A.~J.}\ \bibnamefont {Millar}}, \bibinfo {author} {\bibfnamefont {C.~A.}\ \bibnamefont {O’Hare}},\ and\ \bibinfo {author} {\bibfnamefont {E.}~\bibnamefont {Vitagliano}},\ }\bibfield  {title} {\bibinfo {title} {Dark photon limits: A handbook},\ }\href@noop {} {\bibfield  {journal} {\bibinfo  {journal} {Physical Review D}\ }\textbf {\bibinfo {volume} {104}},\ \bibinfo {pages} {095029} (\bibinfo {year} {2021})}\BibitemShut {NoStop}%
\bibitem [{\citenamefont {Bethe}\ and\ \citenamefont {Salpeter}(2012)}]{bethe2012quantum}%
  \BibitemOpen
  \bibfield  {author} {\bibinfo {author} {\bibfnamefont {H.~A.}\ \bibnamefont {Bethe}}\ and\ \bibinfo {author} {\bibfnamefont {E.~E.}\ \bibnamefont {Salpeter}},\ }\href@noop {} {\emph {\bibinfo {title} {Quantum mechanics of one-and two-electron atoms}}}\ (\bibinfo  {publisher} {Springer Science \& Business Media},\ \bibinfo {year} {2012})\ Chap.~\bibinfo {chapter} {12}\BibitemShut {NoStop}%
\bibitem [{\citenamefont {Kosteleck{\`y}}\ and\ \citenamefont {Vargas}(2015)}]{kostelecky2015lorentz}%
  \BibitemOpen
  \bibfield  {author} {\bibinfo {author} {\bibfnamefont {V.~A.}\ \bibnamefont {Kosteleck{\`y}}}\ and\ \bibinfo {author} {\bibfnamefont {A.~J.}\ \bibnamefont {Vargas}},\ }\bibfield  {title} {\bibinfo {title} {Lorentz and {CPT} tests with hydrogen, antihydrogen, and related systems},\ }\href@noop {} {\bibfield  {journal} {\bibinfo  {journal} {Physical Review D}\ }\textbf {\bibinfo {volume} {92}},\ \bibinfo {pages} {056002} (\bibinfo {year} {2015})}\BibitemShut {NoStop}%
\bibitem [{\citenamefont {Dobrescu}\ and\ \citenamefont {Mocioiu}(2006)}]{Dobrescu:2006au}%
  \BibitemOpen
  \bibfield  {author} {\bibinfo {author} {\bibfnamefont {B.~A.}\ \bibnamefont {Dobrescu}}\ and\ \bibinfo {author} {\bibfnamefont {I.}~\bibnamefont {Mocioiu}},\ }\bibfield  {title} {\bibinfo {title} {{Spin-dependent macroscopic forces from new particle exchange}},\ }\href {https://doi.org/10.1088/1126-6708/2006/11/005} {\bibfield  {journal} {\bibinfo  {journal} {JHEP}\ }\textbf {\bibinfo {volume} {11}},\ \bibinfo {pages} {005}},\ \Eprint {https://arxiv.org/abs/hep-ph/0605342} {arXiv:hep-ph/0605342} \BibitemShut {NoStop}%
\bibitem [{\citenamefont {Cong}\ \emph {et~al.}(2025)\citenamefont {Cong}, \citenamefont {Ji}, \citenamefont {Fadeev}, \citenamefont {Ficek}, \citenamefont {Jiang}, \citenamefont {Flambaum}, \citenamefont {Guan}, \citenamefont {Jackson~Kimball}, \citenamefont {Kozlov}, \citenamefont {Stadnik} \emph {et~al.}}]{cong2025spin}%
  \BibitemOpen
  \bibfield  {author} {\bibinfo {author} {\bibfnamefont {L.}~\bibnamefont {Cong}}, \bibinfo {author} {\bibfnamefont {W.}~\bibnamefont {Ji}}, \bibinfo {author} {\bibfnamefont {P.}~\bibnamefont {Fadeev}}, \bibinfo {author} {\bibfnamefont {F.}~\bibnamefont {Ficek}}, \bibinfo {author} {\bibfnamefont {M.}~\bibnamefont {Jiang}}, \bibinfo {author} {\bibfnamefont {V.~V.}\ \bibnamefont {Flambaum}}, \bibinfo {author} {\bibfnamefont {H.}~\bibnamefont {Guan}}, \bibinfo {author} {\bibfnamefont {D.~F.}\ \bibnamefont {Jackson~Kimball}}, \bibinfo {author} {\bibfnamefont {M.~G.}\ \bibnamefont {Kozlov}}, \bibinfo {author} {\bibfnamefont {Y.~V.}\ \bibnamefont {Stadnik}}, \emph {et~al.},\ }\bibfield  {title} {\bibinfo {title} {Spin-dependent exotic interactions},\ }\href@noop {} {\bibfield  {journal} {\bibinfo  {journal} {Reviews of Modern Physics}\ }\textbf {\bibinfo {volume} {97}},\ \bibinfo {pages} {025005} (\bibinfo {year} {2025})}\BibitemShut {NoStop}%
\bibitem [{\citenamefont {Terrano}\ \emph {et~al.}(2015)\citenamefont {Terrano}, \citenamefont {Adelberger}, \citenamefont {Lee},\ and\ \citenamefont {Heckel}}]{terrano2015short}%
  \BibitemOpen
  \bibfield  {author} {\bibinfo {author} {\bibfnamefont {W.}~\bibnamefont {Terrano}}, \bibinfo {author} {\bibfnamefont {E.}~\bibnamefont {Adelberger}}, \bibinfo {author} {\bibfnamefont {J.}~\bibnamefont {Lee}},\ and\ \bibinfo {author} {\bibfnamefont {B.}~\bibnamefont {Heckel}},\ }\bibfield  {title} {\bibinfo {title} {Short-range, spin-dependent interactions of electrons: A probe for exotic pseudo-goldstone bosons},\ }\href@noop {} {\bibfield  {journal} {\bibinfo  {journal} {Physical review letters}\ }\textbf {\bibinfo {volume} {115}},\ \bibinfo {pages} {201801} (\bibinfo {year} {2015})}\BibitemShut {NoStop}%
\bibitem [{\citenamefont {Graham}\ and\ \citenamefont {Rajendran}(2013)}]{graham2013new}%
  \BibitemOpen
  \bibfield  {author} {\bibinfo {author} {\bibfnamefont {P.~W.}\ \bibnamefont {Graham}}\ and\ \bibinfo {author} {\bibfnamefont {S.}~\bibnamefont {Rajendran}},\ }\bibfield  {title} {\bibinfo {title} {New observables for direct detection of axion dark matter},\ }\href@noop {} {\bibfield  {journal} {\bibinfo  {journal} {Physical Review D—Particles, Fields, Gravitation, and Cosmology}\ }\textbf {\bibinfo {volume} {88}},\ \bibinfo {pages} {035023} (\bibinfo {year} {2013})}\BibitemShut {NoStop}%
\bibitem [{\citenamefont {Antypas}\ \emph {et~al.}(2022)\citenamefont {Antypas}, \citenamefont {Banerjee}, \citenamefont {Bartram}, \citenamefont {Baryakhtar}, \citenamefont {Betz}, \citenamefont {Bollinger}, \citenamefont {Boutan}, \citenamefont {Bowring}, \citenamefont {Budker}, \citenamefont {Carney} \emph {et~al.}}]{antypas2022new}%
  \BibitemOpen
  \bibfield  {author} {\bibinfo {author} {\bibfnamefont {D.}~\bibnamefont {Antypas}}, \bibinfo {author} {\bibfnamefont {A.}~\bibnamefont {Banerjee}}, \bibinfo {author} {\bibfnamefont {C.}~\bibnamefont {Bartram}}, \bibinfo {author} {\bibfnamefont {M.}~\bibnamefont {Baryakhtar}}, \bibinfo {author} {\bibfnamefont {J.}~\bibnamefont {Betz}}, \bibinfo {author} {\bibfnamefont {J.}~\bibnamefont {Bollinger}}, \bibinfo {author} {\bibfnamefont {C.}~\bibnamefont {Boutan}}, \bibinfo {author} {\bibfnamefont {D.}~\bibnamefont {Bowring}}, \bibinfo {author} {\bibfnamefont {D.}~\bibnamefont {Budker}}, \bibinfo {author} {\bibfnamefont {D.}~\bibnamefont {Carney}}, \emph {et~al.},\ }\bibfield  {title} {\bibinfo {title} {New horizons: Scalar and vector ultralight dark matter},\ }\href@noop {} {\bibfield  {journal} {\bibinfo  {journal} {arXiv preprint arXiv:2203.14915}\ } (\bibinfo {year} {2022})}\BibitemShut {NoStop}%
\bibitem [{\citenamefont {Bouchiat}\ and\ \citenamefont {Bouchiat}(1974)}]{bouchiat1974weak}%
  \BibitemOpen
  \bibfield  {author} {\bibinfo {author} {\bibfnamefont {M.}~\bibnamefont {Bouchiat}}\ and\ \bibinfo {author} {\bibfnamefont {C.}~\bibnamefont {Bouchiat}},\ }\bibfield  {title} {\bibinfo {title} {Weak neutral currents in atomic physics},\ }\href@noop {} {\bibfield  {journal} {\bibinfo  {journal} {Physics Letters B}\ }\textbf {\bibinfo {volume} {48}},\ \bibinfo {pages} {111} (\bibinfo {year} {1974})}\BibitemShut {NoStop}%
\bibitem [{\citenamefont {Khriplovich}(1991)}]{khriplovich1991parity}%
  \BibitemOpen
  \bibfield  {author} {\bibinfo {author} {\bibfnamefont {I.~B.}\ \bibnamefont {Khriplovich}},\ }\href@noop {} {\emph {\bibinfo {title} {Parity nonconservation in atomic phenomena}}}\ (\bibinfo  {publisher} {Gordon and Breach Science Publishers},\ \bibinfo {year} {1991})\BibitemShut {NoStop}%
\bibitem [{\citenamefont {Afach}\ \emph {et~al.}(2024)\citenamefont {Afach}, \citenamefont {Aybas~Tumturk}, \citenamefont {Bekker}, \citenamefont {Buchler}, \citenamefont {Budker}, \citenamefont {Cervantes}, \citenamefont {Derevianko}, \citenamefont {Eby}, \citenamefont {Figueroa}, \citenamefont {Folman} \emph {et~al.}}]{afach2024can}%
  \BibitemOpen
  \bibfield  {author} {\bibinfo {author} {\bibfnamefont {S.}~\bibnamefont {Afach}}, \bibinfo {author} {\bibfnamefont {D.}~\bibnamefont {Aybas~Tumturk}}, \bibinfo {author} {\bibfnamefont {H.}~\bibnamefont {Bekker}}, \bibinfo {author} {\bibfnamefont {B.~C.}\ \bibnamefont {Buchler}}, \bibinfo {author} {\bibfnamefont {D.}~\bibnamefont {Budker}}, \bibinfo {author} {\bibfnamefont {K.}~\bibnamefont {Cervantes}}, \bibinfo {author} {\bibfnamefont {A.}~\bibnamefont {Derevianko}}, \bibinfo {author} {\bibfnamefont {J.}~\bibnamefont {Eby}}, \bibinfo {author} {\bibfnamefont {N.~L.}\ \bibnamefont {Figueroa}}, \bibinfo {author} {\bibfnamefont {R.}~\bibnamefont {Folman}}, \emph {et~al.},\ }\bibfield  {title} {\bibinfo {title} {What can a gnome do? search targets for the global network of optical magnetometers for exotic physics searches},\ }\href@noop {} {\bibfield  {journal} {\bibinfo  {journal} {Annalen der Physik}\ }\textbf {\bibinfo {volume} {536}},\ \bibinfo {pages} {2300083} (\bibinfo {year} {2024})}\BibitemShut
  {NoStop}%
\bibitem [{\citenamefont {Lahs}\ and\ \citenamefont {Comparat}(2024)}]{lahs2024polarizabilities}%
  \BibitemOpen
  \bibfield  {author} {\bibinfo {author} {\bibfnamefont {S.}~\bibnamefont {Lahs}}\ and\ \bibinfo {author} {\bibfnamefont {D.}~\bibnamefont {Comparat}},\ }\bibfield  {title} {\bibinfo {title} {Polarizabilities as probes for p, t, and pt violation},\ }\href@noop {} {\bibfield  {journal} {\bibinfo  {journal} {New Journal of Physics}\ } (\bibinfo {year} {2024})}\BibitemShut {NoStop}%
\bibitem [{\citenamefont {Becher}\ \emph {et~al.}(2018)\citenamefont {Becher}, \citenamefont {Baier}, \citenamefont {Aikawa}, \citenamefont {Lepers}, \citenamefont {Wyart}, \citenamefont {Dulieu},\ and\ \citenamefont {Ferlaino}}]{becher2018anisotropic}%
  \BibitemOpen
  \bibfield  {author} {\bibinfo {author} {\bibfnamefont {J.~H.}\ \bibnamefont {Becher}}, \bibinfo {author} {\bibfnamefont {S.}~\bibnamefont {Baier}}, \bibinfo {author} {\bibfnamefont {K.}~\bibnamefont {Aikawa}}, \bibinfo {author} {\bibfnamefont {M.}~\bibnamefont {Lepers}}, \bibinfo {author} {\bibfnamefont {J.-F.}\ \bibnamefont {Wyart}}, \bibinfo {author} {\bibfnamefont {O.}~\bibnamefont {Dulieu}},\ and\ \bibinfo {author} {\bibfnamefont {F.}~\bibnamefont {Ferlaino}},\ }\bibfield  {title} {\bibinfo {title} {Anisotropic polarizability of erbium atoms},\ }\href@noop {} {\bibfield  {journal} {\bibinfo  {journal} {Physical Review A}\ }\textbf {\bibinfo {volume} {97}},\ \bibinfo {pages} {012509} (\bibinfo {year} {2018})}\BibitemShut {NoStop}%
\bibitem [{\citenamefont {Holloway}\ \emph {et~al.}(2014)\citenamefont {Holloway}, \citenamefont {Gordon}, \citenamefont {Jefferts}, \citenamefont {Schwarzkopf}, \citenamefont {Anderson}, \citenamefont {Miller}, \citenamefont {Thaicharoen},\ and\ \citenamefont {Raithel}}]{holloway2014broadband}%
  \BibitemOpen
  \bibfield  {author} {\bibinfo {author} {\bibfnamefont {C.~L.}\ \bibnamefont {Holloway}}, \bibinfo {author} {\bibfnamefont {J.~A.}\ \bibnamefont {Gordon}}, \bibinfo {author} {\bibfnamefont {S.}~\bibnamefont {Jefferts}}, \bibinfo {author} {\bibfnamefont {A.}~\bibnamefont {Schwarzkopf}}, \bibinfo {author} {\bibfnamefont {D.~A.}\ \bibnamefont {Anderson}}, \bibinfo {author} {\bibfnamefont {S.~A.}\ \bibnamefont {Miller}}, \bibinfo {author} {\bibfnamefont {N.}~\bibnamefont {Thaicharoen}},\ and\ \bibinfo {author} {\bibfnamefont {G.}~\bibnamefont {Raithel}},\ }\bibfield  {title} {\bibinfo {title} {Broadband rydberg atom-based electric-field probe for si-traceable, self-calibrated measurements},\ }\href@noop {} {\bibfield  {journal} {\bibinfo  {journal} {IEEE Transactions on Antennas and Propagation}\ }\textbf {\bibinfo {volume} {62}},\ \bibinfo {pages} {6169} (\bibinfo {year} {2014})}\BibitemShut {NoStop}%
\bibitem [{\citenamefont {Yuan}\ \emph {et~al.}(2023)\citenamefont {Yuan}, \citenamefont {Yang}, \citenamefont {Jing}, \citenamefont {Zhang}, \citenamefont {Jiao}, \citenamefont {Li}, \citenamefont {Zhang}, \citenamefont {Xiao},\ and\ \citenamefont {Jia}}]{yuan2023quantum}%
  \BibitemOpen
  \bibfield  {author} {\bibinfo {author} {\bibfnamefont {J.}~\bibnamefont {Yuan}}, \bibinfo {author} {\bibfnamefont {W.}~\bibnamefont {Yang}}, \bibinfo {author} {\bibfnamefont {M.}~\bibnamefont {Jing}}, \bibinfo {author} {\bibfnamefont {H.}~\bibnamefont {Zhang}}, \bibinfo {author} {\bibfnamefont {Y.}~\bibnamefont {Jiao}}, \bibinfo {author} {\bibfnamefont {W.}~\bibnamefont {Li}}, \bibinfo {author} {\bibfnamefont {L.}~\bibnamefont {Zhang}}, \bibinfo {author} {\bibfnamefont {L.}~\bibnamefont {Xiao}},\ and\ \bibinfo {author} {\bibfnamefont {S.}~\bibnamefont {Jia}},\ }\bibfield  {title} {\bibinfo {title} {Quantum sensing of microwave electric fields based on rydberg atoms},\ }\href@noop {} {\bibfield  {journal} {\bibinfo  {journal} {Reports on Progress in Physics}\ }\textbf {\bibinfo {volume} {86}},\ \bibinfo {pages} {106001} (\bibinfo {year} {2023})}\BibitemShut {NoStop}%
\bibitem [{\citenamefont {Matsuki}\ and\ \citenamefont {Yamamoto}(1991)}]{matsuki1991direct}%
  \BibitemOpen
  \bibfield  {author} {\bibinfo {author} {\bibfnamefont {S.}~\bibnamefont {Matsuki}}\ and\ \bibinfo {author} {\bibfnamefont {K.}~\bibnamefont {Yamamoto}},\ }\bibfield  {title} {\bibinfo {title} {Direct detection of galactic axions with rydberg atoms in an inhibited cavity regime},\ }\href@noop {} {\bibfield  {journal} {\bibinfo  {journal} {Physics Letters B}\ }\textbf {\bibinfo {volume} {263}},\ \bibinfo {pages} {523} (\bibinfo {year} {1991})}\BibitemShut {NoStop}%
\bibitem [{\citenamefont {Gu{\'e}}\ \emph {et~al.}(2023)\citenamefont {Gu{\'e}}, \citenamefont {Hees}, \citenamefont {Lodewyck}, \citenamefont {Le~Targat},\ and\ \citenamefont {Wolf}}]{gue2023search}%
  \BibitemOpen
  \bibfield  {author} {\bibinfo {author} {\bibfnamefont {J.}~\bibnamefont {Gu{\'e}}}, \bibinfo {author} {\bibfnamefont {A.}~\bibnamefont {Hees}}, \bibinfo {author} {\bibfnamefont {J.}~\bibnamefont {Lodewyck}}, \bibinfo {author} {\bibfnamefont {R.}~\bibnamefont {Le~Targat}},\ and\ \bibinfo {author} {\bibfnamefont {P.}~\bibnamefont {Wolf}},\ }\bibfield  {title} {\bibinfo {title} {Search for vector dark matter in microwave cavities with rydberg atoms},\ }\href@noop {} {\bibfield  {journal} {\bibinfo  {journal} {Physical Review D}\ }\textbf {\bibinfo {volume} {108}},\ \bibinfo {pages} {035042} (\bibinfo {year} {2023})}\BibitemShut {NoStop}%
\bibitem [{\citenamefont {Abel}\ \emph {et~al.}(2017)\citenamefont {Abel}, \citenamefont {Ayres}, \citenamefont {Ban}, \citenamefont {Bison}, \citenamefont {Bodek}, \citenamefont {Bondar}, \citenamefont {Daum}, \citenamefont {Fairbairn}, \citenamefont {Flambaum}, \citenamefont {Geltenbort} \emph {et~al.}}]{abel2017search}%
  \BibitemOpen
  \bibfield  {author} {\bibinfo {author} {\bibfnamefont {C.}~\bibnamefont {Abel}}, \bibinfo {author} {\bibfnamefont {N.~J.}\ \bibnamefont {Ayres}}, \bibinfo {author} {\bibfnamefont {G.}~\bibnamefont {Ban}}, \bibinfo {author} {\bibfnamefont {G.}~\bibnamefont {Bison}}, \bibinfo {author} {\bibfnamefont {K.}~\bibnamefont {Bodek}}, \bibinfo {author} {\bibfnamefont {V.}~\bibnamefont {Bondar}}, \bibinfo {author} {\bibfnamefont {M.}~\bibnamefont {Daum}}, \bibinfo {author} {\bibfnamefont {M.}~\bibnamefont {Fairbairn}}, \bibinfo {author} {\bibfnamefont {V.~V.}\ \bibnamefont {Flambaum}}, \bibinfo {author} {\bibfnamefont {P.}~\bibnamefont {Geltenbort}}, \emph {et~al.},\ }\bibfield  {title} {\bibinfo {title} {Search for axionlike dark matter through nuclear spin precession in electric and magnetic fields},\ }\href@noop {} {\bibfield  {journal} {\bibinfo  {journal} {Physical Review X}\ }\textbf {\bibinfo {volume} {7}},\ \bibinfo {pages} {041034} (\bibinfo {year} {2017})}\BibitemShut {NoStop}%
\bibitem [{\citenamefont {Roussy}\ \emph {et~al.}(2021)\citenamefont {Roussy}, \citenamefont {Palken}, \citenamefont {Cairncross}, \citenamefont {Brubaker}, \citenamefont {Gresh}, \citenamefont {Grau}, \citenamefont {Cossel}, \citenamefont {Ng}, \citenamefont {Shagam}, \citenamefont {Zhou} \emph {et~al.}}]{roussy2021experimental}%
  \BibitemOpen
  \bibfield  {author} {\bibinfo {author} {\bibfnamefont {T.~S.}\ \bibnamefont {Roussy}}, \bibinfo {author} {\bibfnamefont {D.~A.}\ \bibnamefont {Palken}}, \bibinfo {author} {\bibfnamefont {W.~B.}\ \bibnamefont {Cairncross}}, \bibinfo {author} {\bibfnamefont {B.~M.}\ \bibnamefont {Brubaker}}, \bibinfo {author} {\bibfnamefont {D.~N.}\ \bibnamefont {Gresh}}, \bibinfo {author} {\bibfnamefont {M.}~\bibnamefont {Grau}}, \bibinfo {author} {\bibfnamefont {K.~C.}\ \bibnamefont {Cossel}}, \bibinfo {author} {\bibfnamefont {K.~B.}\ \bibnamefont {Ng}}, \bibinfo {author} {\bibfnamefont {Y.}~\bibnamefont {Shagam}}, \bibinfo {author} {\bibfnamefont {Y.}~\bibnamefont {Zhou}}, \emph {et~al.},\ }\bibfield  {title} {\bibinfo {title} {Experimental constraint on axionlike particles over seven orders of magnitude in mass},\ }\href@noop {} {\bibfield  {journal} {\bibinfo  {journal} {Physical Review Letters}\ }\textbf {\bibinfo {volume} {126}},\ \bibinfo {pages} {171301} (\bibinfo {year} {2021})}\BibitemShut {NoStop}%
\bibitem [{\citenamefont {Budker}\ \emph {et~al.}(2014)\citenamefont {Budker}, \citenamefont {Graham}, \citenamefont {Ledbetter}, \citenamefont {Rajendran},\ and\ \citenamefont {Sushkov}}]{budker2014proposal}%
  \BibitemOpen
  \bibfield  {author} {\bibinfo {author} {\bibfnamefont {D.}~\bibnamefont {Budker}}, \bibinfo {author} {\bibfnamefont {P.~W.}\ \bibnamefont {Graham}}, \bibinfo {author} {\bibfnamefont {M.}~\bibnamefont {Ledbetter}}, \bibinfo {author} {\bibfnamefont {S.}~\bibnamefont {Rajendran}},\ and\ \bibinfo {author} {\bibfnamefont {A.~O.}\ \bibnamefont {Sushkov}},\ }\bibfield  {title} {\bibinfo {title} {Proposal for a cosmic axion spin precession experiment (casper)},\ }\href@noop {} {\bibfield  {journal} {\bibinfo  {journal} {Physical Review X}\ }\textbf {\bibinfo {volume} {4}},\ \bibinfo {pages} {021030} (\bibinfo {year} {2014})}\BibitemShut {NoStop}%
\bibitem [{\citenamefont {Arvanitaki}\ \emph {et~al.}(2024)\citenamefont {Arvanitaki}, \citenamefont {Madden},\ and\ \citenamefont {Van~Tilburg}}]{arvanitaki2024piezoaxionic}%
  \BibitemOpen
  \bibfield  {author} {\bibinfo {author} {\bibfnamefont {A.}~\bibnamefont {Arvanitaki}}, \bibinfo {author} {\bibfnamefont {A.}~\bibnamefont {Madden}},\ and\ \bibinfo {author} {\bibfnamefont {K.}~\bibnamefont {Van~Tilburg}},\ }\bibfield  {title} {\bibinfo {title} {Piezoaxionic effect},\ }\href@noop {} {\bibfield  {journal} {\bibinfo  {journal} {Physical Review D}\ }\textbf {\bibinfo {volume} {109}},\ \bibinfo {pages} {072009} (\bibinfo {year} {2024})}\BibitemShut {NoStop}%
\bibitem [{\citenamefont {Bloch}\ \emph {et~al.}(2023)\citenamefont {Bloch}, \citenamefont {Budker}, \citenamefont {Flambaum}, \citenamefont {Samsonov}, \citenamefont {Sushkov},\ and\ \citenamefont {Tretiak}}]{bloch2023scalar}%
  \BibitemOpen
  \bibfield  {author} {\bibinfo {author} {\bibfnamefont {I.}~\bibnamefont {Bloch}}, \bibinfo {author} {\bibfnamefont {D.}~\bibnamefont {Budker}}, \bibinfo {author} {\bibfnamefont {V.}~\bibnamefont {Flambaum}}, \bibinfo {author} {\bibfnamefont {I.}~\bibnamefont {Samsonov}}, \bibinfo {author} {\bibfnamefont {A.}~\bibnamefont {Sushkov}},\ and\ \bibinfo {author} {\bibfnamefont {O.}~\bibnamefont {Tretiak}},\ }\bibfield  {title} {\bibinfo {title} {Scalar dark matter induced oscillation of a permanent-magnet field},\ }\href@noop {} {\bibfield  {journal} {\bibinfo  {journal} {Physical Review D}\ }\textbf {\bibinfo {volume} {107}},\ \bibinfo {pages} {075033} (\bibinfo {year} {2023})}\BibitemShut {NoStop}%
\bibitem [{\citenamefont {Nanda}\ \emph {et~al.}(2025)\citenamefont {Nanda}, \citenamefont {Comparat}, \citenamefont {Dulieu}, \citenamefont {Lahs}, \citenamefont {Malbrunot}, \citenamefont {Nowak}, \citenamefont {Simon},\ and\ \citenamefont {Widmann}}]{nanda2025demonstration}%
  \BibitemOpen
  \bibfield  {author} {\bibinfo {author} {\bibfnamefont {A.}~\bibnamefont {Nanda}}, \bibinfo {author} {\bibfnamefont {D.}~\bibnamefont {Comparat}}, \bibinfo {author} {\bibfnamefont {O.}~\bibnamefont {Dulieu}}, \bibinfo {author} {\bibfnamefont {S.}~\bibnamefont {Lahs}}, \bibinfo {author} {\bibfnamefont {C.}~\bibnamefont {Malbrunot}}, \bibinfo {author} {\bibfnamefont {L.}~\bibnamefont {Nowak}}, \bibinfo {author} {\bibfnamefont {M.~C.}\ \bibnamefont {Simon}},\ and\ \bibinfo {author} {\bibfnamefont {E.}~\bibnamefont {Widmann}},\ }\bibfield  {title} {\bibinfo {title} {Demonstration of deuterium's enhanced sensitivity to symmetry violations governed by the standard-model extension},\ }\href@noop {} {\bibfield  {journal} {\bibinfo  {journal} {arXiv preprint arXiv:2507.07473}\ } (\bibinfo {year} {2025})}\BibitemShut {NoStop}%
\bibitem [{\citenamefont {Humphrey}\ \emph {et~al.}(2003)\citenamefont {Humphrey}, \citenamefont {Phillips}, \citenamefont {Mattison}, \citenamefont {Vessot}, \citenamefont {Stoner},\ and\ \citenamefont {Walsworth}}]{humphrey2003testing}%
  \BibitemOpen
  \bibfield  {author} {\bibinfo {author} {\bibfnamefont {M.~A.}\ \bibnamefont {Humphrey}}, \bibinfo {author} {\bibfnamefont {D.~F.}\ \bibnamefont {Phillips}}, \bibinfo {author} {\bibfnamefont {E.~M.}\ \bibnamefont {Mattison}}, \bibinfo {author} {\bibfnamefont {R.~F.}\ \bibnamefont {Vessot}}, \bibinfo {author} {\bibfnamefont {R.~E.}\ \bibnamefont {Stoner}},\ and\ \bibinfo {author} {\bibfnamefont {R.~L.}\ \bibnamefont {Walsworth}},\ }\bibfield  {title} {\bibinfo {title} {Testing {CPT} and lorentz symmetry with hydrogen masers},\ }\href@noop {} {\bibfield  {journal} {\bibinfo  {journal} {Physical Review A}\ }\textbf {\bibinfo {volume} {68}},\ \bibinfo {pages} {063807} (\bibinfo {year} {2003})}\BibitemShut {NoStop}%
\bibitem [{\citenamefont {Nowak}\ \emph {et~al.}(2024)\citenamefont {Nowak}, \citenamefont {Malbrunot}, \citenamefont {Simon}, \citenamefont {Amsler}, \citenamefont {Cuendis}, \citenamefont {Lahs}, \citenamefont {Lanz}, \citenamefont {Nanda}, \citenamefont {Wiesinger}, \citenamefont {Wolz} \emph {et~al.}}]{nowak2024cpt}%
  \BibitemOpen
  \bibfield  {author} {\bibinfo {author} {\bibfnamefont {L.}~\bibnamefont {Nowak}}, \bibinfo {author} {\bibfnamefont {C.}~\bibnamefont {Malbrunot}}, \bibinfo {author} {\bibfnamefont {M.~C.}\ \bibnamefont {Simon}}, \bibinfo {author} {\bibfnamefont {C.}~\bibnamefont {Amsler}}, \bibinfo {author} {\bibfnamefont {S.~A.}\ \bibnamefont {Cuendis}}, \bibinfo {author} {\bibfnamefont {S.}~\bibnamefont {Lahs}}, \bibinfo {author} {\bibfnamefont {A.}~\bibnamefont {Lanz}}, \bibinfo {author} {\bibfnamefont {A.}~\bibnamefont {Nanda}}, \bibinfo {author} {\bibfnamefont {M.}~\bibnamefont {Wiesinger}}, \bibinfo {author} {\bibfnamefont {T.}~\bibnamefont {Wolz}}, \emph {et~al.},\ }\bibfield  {title} {\bibinfo {title} {{CPT} and lorentz symmetry tests with hydrogen using a novel in-beam hyperfine spectroscopy method applicable to antihydrogen experiments},\ }\href@noop {} {\bibfield  {journal} {\bibinfo  {journal} {Physics Letters B}\ }\textbf {\bibinfo {volume} {858}},\ \bibinfo {pages} {139012} (\bibinfo {year}
  {2024})}\BibitemShut {NoStop}%
\bibitem [{\citenamefont {Schiff}(1963)}]{schiff1963measurability}%
  \BibitemOpen
  \bibfield  {author} {\bibinfo {author} {\bibfnamefont {L.}~\bibnamefont {Schiff}},\ }\bibfield  {title} {\bibinfo {title} {Measurability of nuclear electric dipole moments},\ }\href@noop {} {\bibfield  {journal} {\bibinfo  {journal} {Phys. Rev.}\ }\textbf {\bibinfo {volume} {132}},\ \bibinfo {pages} {2194} (\bibinfo {year} {1963})}\BibitemShut {NoStop}%
\bibitem [{\citenamefont {Sandars}(1965)}]{sandars1965electric}%
  \BibitemOpen
  \bibfield  {author} {\bibinfo {author} {\bibfnamefont {P.~G.~H.}\ \bibnamefont {Sandars}},\ }\bibfield  {title} {\bibinfo {title} {The electric dipole moment of an atom},\ }\href@noop {} {\bibfield  {journal} {\bibinfo  {journal} {Phys. Lett.}\ }\textbf {\bibinfo {volume} {14}},\ \bibinfo {pages} {194} (\bibinfo {year} {1965})}\BibitemShut {NoStop}%
\bibitem [{\citenamefont {Flambaum}(2023)}]{flambaum2023screening}%
  \BibitemOpen
  \bibfield  {author} {\bibinfo {author} {\bibfnamefont {V.}~\bibnamefont {Flambaum}},\ }\bibfield  {title} {\bibinfo {title} {Screening of the electric field and nuclear electric dipole moment in nonstationary states of atoms and molecules},\ }\href@noop {} {\bibfield  {journal} {\bibinfo  {journal} {Physical Review A}\ }\textbf {\bibinfo {volume} {108}},\ \bibinfo {pages} {L030801} (\bibinfo {year} {2023})}\BibitemShut {NoStop}%
\bibitem [{\citenamefont {Ginges}\ and\ \citenamefont {Flambaum}(2004)}]{ginges2004violations}%
  \BibitemOpen
  \bibfield  {author} {\bibinfo {author} {\bibfnamefont {J.}~\bibnamefont {Ginges}}\ and\ \bibinfo {author} {\bibfnamefont {V.~V.}\ \bibnamefont {Flambaum}},\ }\bibfield  {title} {\bibinfo {title} {Violations of fundamental symmetries in atoms and tests of unification theories of elementary particles},\ }\href@noop {} {\bibfield  {journal} {\bibinfo  {journal} {Physics Reports}\ }\textbf {\bibinfo {volume} {397}},\ \bibinfo {pages} {63} (\bibinfo {year} {2004})}\BibitemShut {NoStop}%
\bibitem [{\citenamefont {Guena}\ \emph {et~al.}(2005)\citenamefont {Guena}, \citenamefont {Lintz},\ and\ \citenamefont {Bouchiat}}]{guena2005atomic}%
  \BibitemOpen
  \bibfield  {author} {\bibinfo {author} {\bibfnamefont {J.}~\bibnamefont {Guena}}, \bibinfo {author} {\bibfnamefont {M.}~\bibnamefont {Lintz}},\ and\ \bibinfo {author} {\bibfnamefont {M.}~\bibnamefont {Bouchiat}},\ }\bibfield  {title} {\bibinfo {title} {Atomic parity violation: Principles, recent results, present motivations},\ }\href@noop {} {\bibfield  {journal} {\bibinfo  {journal} {Modern Physics Letters A}\ }\textbf {\bibinfo {volume} {20}},\ \bibinfo {pages} {375} (\bibinfo {year} {2005})}\BibitemShut {NoStop}%
\bibitem [{\citenamefont {Wood}\ \emph {et~al.}(1997)\citenamefont {Wood}, \citenamefont {Bennett}, \citenamefont {Cho}, \citenamefont {Masterson}, \citenamefont {Roberts}, \citenamefont {Tanner},\ and\ \citenamefont {Wieman}}]{wood1997measurement}%
  \BibitemOpen
  \bibfield  {author} {\bibinfo {author} {\bibfnamefont {C.}~\bibnamefont {Wood}}, \bibinfo {author} {\bibfnamefont {S.}~\bibnamefont {Bennett}}, \bibinfo {author} {\bibfnamefont {D.}~\bibnamefont {Cho}}, \bibinfo {author} {\bibfnamefont {B.}~\bibnamefont {Masterson}}, \bibinfo {author} {\bibfnamefont {J.}~\bibnamefont {Roberts}}, \bibinfo {author} {\bibfnamefont {C.}~\bibnamefont {Tanner}},\ and\ \bibinfo {author} {\bibfnamefont {C.~E.}\ \bibnamefont {Wieman}},\ }\bibfield  {title} {\bibinfo {title} {Measurement of parity nonconservation and an anapole moment in cesium},\ }\href@noop {} {\bibfield  {journal} {\bibinfo  {journal} {Science}\ }\textbf {\bibinfo {volume} {275}},\ \bibinfo {pages} {1759} (\bibinfo {year} {1997})}\BibitemShut {NoStop}%
\bibitem [{\citenamefont {Langhoff}\ \emph {et~al.}(1972)\citenamefont {Langhoff}, \citenamefont {Epstein},\ and\ \citenamefont {Karplus}}]{langhoff1972aspects}%
  \BibitemOpen
  \bibfield  {author} {\bibinfo {author} {\bibfnamefont {P.}~\bibnamefont {Langhoff}}, \bibinfo {author} {\bibfnamefont {S.}~\bibnamefont {Epstein}},\ and\ \bibinfo {author} {\bibfnamefont {M.}~\bibnamefont {Karplus}},\ }\bibfield  {title} {\bibinfo {title} {Aspects of time-dependent perturbation theory},\ }\href@noop {} {\bibfield  {journal} {\bibinfo  {journal} {Reviews of Modern Physics}\ }\textbf {\bibinfo {volume} {44}},\ \bibinfo {pages} {602} (\bibinfo {year} {1972})}\BibitemShut {NoStop}%
\bibitem [{\citenamefont {Landau}\ and\ \citenamefont {Lishitz}(2013)}]{landau2013quantum6}%
  \BibitemOpen
  \bibfield  {author} {\bibinfo {author} {\bibfnamefont {L.}~\bibnamefont {Landau}}\ and\ \bibinfo {author} {\bibfnamefont {E.}~\bibnamefont {Lishitz}},\ }\href@noop {} {\emph {\bibinfo {title} {Quantum mechanics: non-relativistic theory}}},\ Vol.~\bibinfo {volume} {3}\ (\bibinfo  {publisher} {Elsevier},\ \bibinfo {year} {2013})\ Chap.~\bibinfo {chapter} {6}\BibitemShut {NoStop}%
\bibitem [{\citenamefont {Vexiau}\ \emph {et~al.}(2017)\citenamefont {Vexiau}, \citenamefont {Borsalino}, \citenamefont {Lepers}, \citenamefont {Orb{\'a}n}, \citenamefont {Aymar}, \citenamefont {Dulieu},\ and\ \citenamefont {Bouloufa-Maafa}}]{vexiau2017dynamic}%
  \BibitemOpen
  \bibfield  {author} {\bibinfo {author} {\bibfnamefont {R.}~\bibnamefont {Vexiau}}, \bibinfo {author} {\bibfnamefont {D.}~\bibnamefont {Borsalino}}, \bibinfo {author} {\bibfnamefont {M.}~\bibnamefont {Lepers}}, \bibinfo {author} {\bibfnamefont {A.}~\bibnamefont {Orb{\'a}n}}, \bibinfo {author} {\bibfnamefont {M.}~\bibnamefont {Aymar}}, \bibinfo {author} {\bibfnamefont {O.}~\bibnamefont {Dulieu}},\ and\ \bibinfo {author} {\bibfnamefont {N.}~\bibnamefont {Bouloufa-Maafa}},\ }\bibfield  {title} {\bibinfo {title} {Dynamic dipole polarizabilities of heteronuclear alkali dimers: optical response, trapping and control of ultracold molecules},\ }\href@noop {} {\bibfield  {journal} {\bibinfo  {journal} {International Reviews in Physical Chemistry}\ }\textbf {\bibinfo {volume} {36}},\ \bibinfo {pages} {709} (\bibinfo {year} {2017})}\BibitemShut {NoStop}%
\bibitem [{\citenamefont {Varshalovich}\ \emph {et~al.}(1988)\citenamefont {Varshalovich}, \citenamefont {Moskalev},\ and\ \citenamefont {Khersonskii}}]{varshalovich1988quantum}%
  \BibitemOpen
  \bibfield  {author} {\bibinfo {author} {\bibfnamefont {D.~A.}\ \bibnamefont {Varshalovich}}, \bibinfo {author} {\bibfnamefont {A.~N.}\ \bibnamefont {Moskalev}},\ and\ \bibinfo {author} {\bibfnamefont {V.~K.}\ \bibnamefont {Khersonskii}},\ }\href@noop {} {\emph {\bibinfo {title} {Quantum theory of angular momentum}}}\ (\bibinfo  {publisher} {World Scientific},\ \bibinfo {year} {1988})\ Chap.\ \bibinfo {chapter} {3.2}\BibitemShut {NoStop}%
\bibitem [{\citenamefont {Stephens}(1993)}]{stephens1993time}%
  \BibitemOpen
  \bibfield  {author} {\bibinfo {author} {\bibfnamefont {E.}~\bibnamefont {Stephens}},\ }\emph {\bibinfo {title} {Time Reversal Violation in Atoms}},\ \href@noop {} {Ph.D. thesis},\ \bibinfo  {school} {University of Oxford} (\bibinfo {year} {1993})\BibitemShut {NoStop}%
\bibitem [{\citenamefont {Itin}\ and\ \citenamefont {Reches}(2022)}]{itin2022decomposition}%
  \BibitemOpen
  \bibfield  {author} {\bibinfo {author} {\bibfnamefont {Y.}~\bibnamefont {Itin}}\ and\ \bibinfo {author} {\bibfnamefont {S.}~\bibnamefont {Reches}},\ }\bibfield  {title} {\bibinfo {title} {Decomposition of third-order constitutive tensors},\ }\href@noop {} {\bibfield  {journal} {\bibinfo  {journal} {Mathematics and Mechanics of Solids}\ }\textbf {\bibinfo {volume} {27}},\ \bibinfo {pages} {222} (\bibinfo {year} {2022})}\BibitemShut {NoStop}%
\bibitem [{\citenamefont {Sakurai}(1967)}]{sakurai1967advanced}%
  \BibitemOpen
  \bibfield  {author} {\bibinfo {author} {\bibfnamefont {J.~J.}\ \bibnamefont {Sakurai}},\ }\href@noop {} {\emph {\bibinfo {title} {Advanced quantum mechanics}}}\ (\bibinfo  {publisher} {Pearson Education},\ \bibinfo {year} {1967})\BibitemShut {NoStop}%
\bibitem [{\citenamefont {Quevillon}\ and\ \citenamefont {Smith}(2019)}]{quevillon2019axions}%
  \BibitemOpen
  \bibfield  {author} {\bibinfo {author} {\bibfnamefont {J.}~\bibnamefont {Quevillon}}\ and\ \bibinfo {author} {\bibfnamefont {C.}~\bibnamefont {Smith}},\ }\bibfield  {title} {\bibinfo {title} {Axions are blind to anomalies},\ }\href@noop {} {\bibfield  {journal} {\bibinfo  {journal} {The European Physical Journal C}\ }\textbf {\bibinfo {volume} {79}},\ \bibinfo {pages} {1} (\bibinfo {year} {2019})}\BibitemShut {NoStop}%
\bibitem [{\citenamefont {Roberts}\ and\ \citenamefont {Marciano}(2010)}]{roberts2010lepton}%
  \BibitemOpen
  \bibfield  {author} {\bibinfo {author} {\bibfnamefont {B.~L.}\ \bibnamefont {Roberts}}\ and\ \bibinfo {author} {\bibfnamefont {W.~J.}\ \bibnamefont {Marciano}},\ }\href@noop {} {\emph {\bibinfo {title} {Lepton dipole moments}}},\ Vol.~\bibinfo {volume} {20}\ (\bibinfo  {publisher} {World Scientific},\ \bibinfo {year} {2010})\BibitemShut {NoStop}%
\end{thebibliography}%
\end{document}